\documentclass[twocolumn]{aastex63}

\usepackage[utf8]{inputenc}

\usepackage{natbib}
\usepackage{graphicx}
\usepackage{amsmath}
\usepackage{booktabs}
\usepackage{longtable}
\usepackage{xspace}
\usepackage{hyperref}
\usepackage[T1]{fontenc}
\usepackage{lipsum}
\usepackage{comment}
\usepackage{xcolor}
\usepackage{multirow}

\usepackage{IEEEtrantools}
\usepackage{amsmath}

\newcommand{\ms}{ms^{-1}}

\providecommand{\mj}{\ensuremath{\,M_{\rm J}}}
\providecommand{\rj}{\ensuremath{\,R_{\rm J}}}
\newcommand{\rearth}{\mbox{$R_{\ensuremath{\oplus}}$}}
\newcommand{\mearth}{\mbox{$M_{\ensuremath{\oplus}}$}}

\newcommand{\tess}{{\it TESS}}
\newcommand{\jw}{{\it JWST}}

\newcommand{\nrvsHIRES}{73}
\newcommand{\nrvsAPF}{104}

\newcommand{\baseline}{415}

\newcommand{\Kb}{\ensuremath{3.8\pm 0.3}}
\newcommand{\Kc}{\ensuremath{1.8\pm 0.3}}
\newcommand{\Kd}{\ensuremath{0.6\pm 0.3}}
\newcommand{\Ke}{\ensuremath{17.2\pm 0.4}}

\newcommand{\mb}{\ensuremath{10.4 \pm 0.9}}
\newcommand{\mc}{\ensuremath{7.2 \pm 1.4}}
\newcommand{\md}{\ensuremath{2.8 \pm 1.5}}
\newcommand{\me}{\ensuremath{0.34 \pm 0.01}}
\newcommand{\meearth}{\ensuremath{108 \pm 3} }

\newcommand{\DGperL}{\ensuremath{1700}}
\newcommand{\DGperH}{\ensuremath{7200}}
\newcommand{\DGmassL}{\ensuremath{2}}
\newcommand{\DGmassH}{\ensuremath{11}}
\newcommand{\DGsemiL}{\ensuremath{2.6}}
\newcommand{\DGsemiH}{\ensuremath{7.0}}

\newcommand{\rhob}{\ensuremath{1.5 \pm 0.2}}
\newcommand{\rhoc}{\ensuremath{1.4 \pm 0.3}}
\newcommand{\rhod}{\ensuremath{0.5 \pm 0.3}}

\newcommand{\rb}{\ensuremath{3.39 \pm 0.07}}
\newcommand{\rc}{\ensuremath{3.08 \pm 0.07}}
\newcommand{\rd}{\ensuremath{3.04 \pm 0.07}}

\newcommand{\atmb}{\ensuremath{6.5 \pm 0.5}}
\newcommand{\atmc}{\ensuremath{5.7 \pm 0.6}}
\newcommand{\atmd}{\ensuremath{6.4 \pm 0.5}}

\newcommand{\gammadot}{\ensuremath{0.114 \pm 0.006}}
\newcommand{\gammaddot}{\ensuremath{(-6 \pm 2)\times 10^{-5}}}

\newcommand{\TSMb}{\ensuremath{151 \pm 18}}
\newcommand{\SNROneTransitb}{\ensuremath{84 \pm 10}}
\newcommand{\TSMc}{\ensuremath{106 \pm 24}}
\newcommand{\SNROneTransitc}{\ensuremath{71 \pm 16}}
\newcommand{\TSMd}{\ensuremath{>72}\xspace}
\newcommand{\SNROneTransitd}{\ensuremath{>53}\xspace}

\begin{document}

\title{TESS-Keck Survey IX: Masses of Three Sub-Neptunes Orbiting HD 191939 and the Discovery of a Warm Jovian Plus a Distant Sub-Stellar Companion}

\author[0000-0001-8342-7736]{Jack Lubin}
\affiliation{Department of Physics \& Astronomy, University of California Irvine, Irvine, CA 92697, USA}

\author[0000-0002-4290-6826]{Judah Van Zandt}
\affiliation{Department of Physics \& Astronomy, University of California Los Angeles, Los Angeles, CA 90095, USA}

\author[0000-0002-5034-9476]{Rae Holcomb}
\affiliation{Department of Physics \& Astronomy, University of California Irvine, Irvine, CA 92697, USA}

\author[0000-0002-3725-3058]{Lauren M. Weiss}
\affiliation{Institute for Astronomy, University of Hawai`i, 2680 Woodlawn Drive, Honolulu, HI 96822, USA}

\author[0000-0003-0967-2893]{Erik A Petigura}
\affiliation{Department of Physics \& Astronomy, University of California Los Angeles, Los Angeles, CA 90095, USA}

\author[0000-0003-0149-9678]{Paul Robertson}
\affiliation{Department of Physics \& Astronomy, University of California Irvine, Irvine, CA 92697, USA}

\author[0000-0001-8898-8284]{Joseph M. Akana Murphy}
\altaffiliation{NSF Graduate Research Fellow}
\affiliation{Department of Astronomy and Astrophysics, University of California, Santa Cruz, CA 95060, USA}

\author[0000-0003-3623-7280]{Nicholas Scarsdale}
\affiliation{Department of Astronomy and Astrophysics, University of California, Santa Cruz, CA 95060, USA}

\author[0000-0002-7094-7908]{Konstantin Batygin}
\affiliation{Division of Geological and Planetary Sciences California Institute of Technology, Pasadena, CA 91125, USA}

\author[0000-0001-7047-8681]{Alex S. Polanski}
\affiliation{Department of Physics \& Astronomy, University of Kansas, 1082 Malott,1251 Wescoe Hall Dr., Lawrence, KS 66045, USA}

\author[0000-0002-7030-9519]{Natalie M. Batalha}
\affiliation{Department of Astronomy and Astrophysics, University of California, Santa Cruz, CA 95060, USA}

\author{Ian J. M. Crossfield}
\affiliation{Department of Physics \& Astronomy, University of Kansas, 1082 Malott,1251 Wescoe Hall Dr., Lawrence, KS 66045, USA}

\author[0000-0001-8189-0233]{Courtney Dressing}
\affiliation{Department of Astronomy, University of California Berkeley, Berkeley CA 94720, USA}

\author[0000-0003-3504-5316]{Benjamin Fulton}
\affiliation{NASA Exoplanet Science Institute/Caltech-IPAC, MC 314-6, 1200 E California Blvd, Pasadena, CA 91125, USA}

\author[0000-0001-8638-0320]{Andrew W. Howard}
\affiliation{Department of Astronomy, California Institute of Technology, Pasadena, CA 91125, USA}

\author[0000-0001-8832-4488]{Daniel Huber}
\affiliation{Institute for Astronomy, University of Hawai`i, 2680 Woodlawn Drive, Honolulu, HI 96822, USA}

\author[0000-0002-0531-1073]{Howard Isaacson}
\affiliation{Department of Astronomy,  University of California Berkeley, Berkeley CA 94720, USA}
\affiliation{Centre for Astrophysics, University of Southern Queensland, Toowoomba, QLD, Australia}

\author[0000-0002-7084-0529]{Stephen R. Kane}
\affiliation{Department of Earth and Planetary Sciences, University of California, Riverside, CA 92521, USA}

\author[0000-0001-8127-5775]{Arpita Roy}
\affiliation{Space Telescope Science Institute, 3700 San Martin Drive, Baltimore, MD 21218, USA}
\affiliation{Department of Physics and Astronomy, Johns Hopkins University, 3400 N Charles St, Baltimore, MD 21218, USA}

\author[0000-0001-7708-2364]{Corey Beard}
\affiliation{Department of Physics \& Astronomy, University of California Irvine, Irvine, CA 92697, USA}

\author[0000-0002-3199-2888]{Sarah Blunt}
\altaffiliation{NSF Graduate Research Fellow}
\affiliation{Department of Astronomy, California Institute of Technology, Pasadena, CA 91125, USA}

\author[0000-0003-1125-2564]{Ashley Chontos}
\altaffiliation{NSF Graduate Research Fellow}
\affiliation{Institute for Astronomy, University of Hawai`i, 2680 Woodlawn Drive, Honolulu, HI 96822, USA}

\author[0000-0002-8958-0683]{Fei Dai}
\affiliation{Division of Geological and Planetary Sciences
1200 E California Blvd, Pasadena, CA, 91125, USA}

\author[0000-0002-4297-5506]{Paul A. Dalba}
\altaffiliation{NSF Astronomy and Astrophysics Postdoctoral Fellow}
\affiliation{Department of Earth and Planetary Sciences, University of California, Riverside, CA 92521, USA}

\author{Kaz Gary}
\affiliation{Department of Physics \& Astronomy, University of Kansas, 1082 Malott,1251 Wescoe Hall Dr., Lawrence, KS 66045, USA}

\author[0000-0002-8965-3969]{Steven Giacalone}
\affil{Department of Astronomy, University of California Berkeley, Berkeley CA 94720, USA}

\author[0000-0002-0139-4756]{Michelle L. Hill}
\affiliation{Department of Earth and Planetary Sciences, University of California, Riverside, CA 92521, USA}

\author[0000-0002-7216-2135]{Andrew Mayo}
\affil{Department of Astronomy, University of California Berkeley, Berkeley CA 94720, USA}

\author[0000-0003-4603-556X]{Teo Mo\v{c}nik}
\affiliation{Gemini Observatory/NSF's NOIRLab, 670 N. A'ohoku Place, Hilo, HI 96720, USA}

\author[0000-0002-6115-4359]{Molly R. Kosiarek}
\altaffiliation{NSF Graduate Research Fellow}
\affiliation{Department of Astronomy and Astrophysics, University of California, Santa Cruz, CA 95064, USA}

\author[0000-0002-7670-670X]{Malena Rice}
\altaffiliation{NSF Graduate Research Fellow}
\affiliation{Department of Astronomy, Yale University, New Haven, CT 06511, USA}

\author[0000-0003-3856-3143]{Ryan A. Rubenzahl}
\altaffiliation{NSF Graduate Research Fellow}
\affiliation{Department of Astronomy, California Institute of Technology, Pasadena, CA 91125, USA}

\author[0000-0001-9911-7388]{David W. Latham} 
\affil{Center for Astrophysics | Harvard \& Smithsonian, 60 Garden Street, Cambridge, Massachusetts 02138, USA}

\author[0000-0002-6892-6948]{S.~Seager}
\affiliation{Department of Physics and Kavli Institute for Astrophysics and Space Research, Massachusetts Institute of Technology, Cambridge, MA 02139, USA}
\affiliation{Department of Earth, Atmospheric and Planetary Sciences, Massachusetts Institute of Technology, Cambridge, MA 02139, USA}
\affiliation{Department of Aeronautics and Astronautics, MIT, 77 Massachusetts Avenue, Cambridge, MA 02139, USA}

\author[0000-0002-4265-047X]{Joshua N. Winn}
\affiliation{Department of Astrophysical Sciences, Princeton University, 4 Ivy Lane, Princeton, NJ 08544, USA}

\begin{abstract}

Exoplanet systems with multiple transiting planets are natural laboratories for testing planetary astrophysics. One such system is HD 191939 (TOI-1339), a bright (V=9) and Sun-like (G9V) star, which TESS found to host three transiting planets (b, c, and d). The planets have periods of 9, 29, and 38 days each with similar sizes from 3 to 3.4 $\rearth$. To further characterize the system, we measured the radial velocity (RV) of HD 191939 over $\baseline$ days with Keck/HIRES and APF/Levy. We find that $M_b = \mb \mearth$ and $M_c = \mc \mearth$, which are low compared to most known planets of comparable radii. The RVs yield only an upper-limit on $M_d$ (<5.8 $\mearth$ at 2$\sigma$). The RVs further reveal a fourth planet (e) with a minimum mass of $\me \, M_{Jup}$ and an orbital period of 101.4 $\pm$ 0.4 days. Despite its non-transiting geometry, secular interactions between planet e and the inner transiting planets indicate that planet e is coplanar with the transiting planets ($\Delta$\textit{i} < 10$^{\circ}$). We identify a second high mass planet (f) with 95\% confidence intervals on mass between \DGmassL--\DGmassH \, $M_{Jup}$ and period between \DGperL--\DGperH \, days, based on a joint analysis of RVs and astrometry from $Gaia$ and $Hipparcos$. As a bright star hosting multiple planets with well-measured masses, HD 191939 presents many options for comparative planetary astronomy including characterization with JWST.

\keywords{HD 191939, TESS, Keck HIRES, Multi-Planet}
\end{abstract}

\section{Introduction}

\par Bright systems with multiple planets are valuable to the exoplanet community. They are amenable to precise RV monitoring and are natural laboratories of planetary astrophysics. With multiple planets forming from the same protoplanetary disk, such systems allow for comparative exoplanetology investigations, as we can assume a similar history of formation conditions for each planet. 

\par NASA's \textit{Transiting Exoplanet Survey Satellite} ($\tess$; \citealt{ricker2015}) is an all-sky photometric survey searching for planets around the brightest stars, and its discoveries continue to deliver new planetary systems for detailed investigation. Due to the 28-day per sector survey strategy, $\tess$ is finding many exoplanets in short period orbits (<14 days). A primary science goal of the $\tess$ mission is to measure the masses of 50 planets smaller than 4 Earth radii.

\par The TESS-Keck Survey (TKS) is a collaboration among astronomers at Keck partner institutions to combine efforts and telescope time to meet and exceed this science goal (see TKS-0 \citep{Chontos2021}, TKS-I \citep{Dalba2020}, TKS-II \citep{Weiss2021}, TKS-III \citep{Dai2020}, TKS-IV \citep{Rubenzhal2021}). Our survey is further concerned with the formation, evolution, and dynamics of various types of exoplanetary systems. Three of TKS's main goals are characterizing systems with multiple planets, those with possible distant giant planets, and those that show promise for high-quality atmospheric characterization \citep{Chontos2021}

\par HD 191939 is a solar-like star (G9V) that hosts a multi planet system which addresses most of our areas of interest. $\tess$ observed the star for 252 days in 9 non-consecutive sectors during its primary mission, allowing for a long baseline (326 days) of photometry and enabling discovery of longer-period planets. \citet{BA2020} have already announced three transiting planets, two of which would not have been discovered without multiple sectors of coverage. This work includes the first mass measurements of the transiting planets and we have uncovered an additional Jovian planet as well as a high mass planet. In all, this system has a wide diversity of planet masses and periods. 

\par We find that the transiting planets of HD 191939 fall into some of the patterns uncovered by statistics papers on the \textit{Kepler} planets. They have nearly identical radii, as is typical of the Kepler planets \citep{Weiss2018}, yet their spacing is irregular. They have similar masses, consistent with the pattern found in \citet{Millholland2017}, but the planets have low masses for their sizes \citep{Weiss2014}, implying lower than average densities. The masses we present are some of the most precise mass measurements of small transiting planets in a multiplanet system (2 of 3 with 5$\sigma$ mass or better).

\par In this paper, we describe our data sources (\S\ref{datasources}) and analyze the system properties, describing the host star properties (\S\ref{HostStar}) as well as our RV model (\S\ref{ModelComp}) and photometry model (\S\ref{photometrymodel}). Next, we describe the densities and compositions of the transiting planets (\S\ref{bulkcomposition}). We then explore the system dynamics in detail, including constraining the properties of planet f with new techniques (\S\ref{5thcomp}), placing limits on the inclination of planet e (\S\ref{LLsection}), and describing the resonant interactions of planets c and d (\S\ref{mmr}). We then quantify the possibility of additional planets in the system (\S\ref{gapcomplex}) before investigating follow up opportunities for HD 191939 by examining the system's atmospheric and Rossiter-McLaughlin prospects (\S\ref{AtmosphericProspects}). We present our conclusions in \S\ref{conclusion}.

\section{Observations}
\label{datasources}

\subsection{$\tess$ Photometry}

\par Due to the star's high northern declination, $\tess$ observed HD 191939 for a total of 9 sectors in Cycle 2. Data were obtained with a 2-minute cadence during sectors 15-19, 21-22, and 24-25, spanning a total baseline of 326 days from 2019-07-18 to 2020-06-08, though the star was not observed for the entirety of this time \citep{Stassun2018}. We downloaded data processed through the Science Processing Operations Center (SPOC) pipeline through the Mikulski Archive for Space Telescopes (MAST), and used the Pre-search Data Conditioning (PDC) light curves for our analysis. \citep{Jenkins2016}. 

\subsection{Radial Velocities}

\par We acquired $\nrvsHIRES$ RV observations with Keck/HIRES at the W.M. Keck Observatory on Maunakea, Hawaii between November 2019 and December 2020, see Table \ref{tbl:RVData}. We reduced the spectra in the standard procedure of the California Planet Search \citep{Howard2010}. We used a high SNR template from Keck/HIRES to generate a deconvolved stellar spectral template (DSST). We took all RV observations with a warm iodine cell in the light path for wavelength calibration \citep{Valenti1995, Butler1996} with median SNR of $\sim$216 per pixel at the iodine wavelength region of $\sim$500 nm. 

\par We also acquired $\nrvsAPF$ RV observations with the Automated Planet Finder telescope (APF) \citep{Vogt2014} at Lick Observatory in California between December 2019 and December 2020. At the beginning of the baseline, we observed twice per night and binned the two observations. After February 2020, we changed our observing strategy to obtain one spectrum per night due to time constraints within our survey. We used the same Keck/HIRES template to calculate the APF RVs because it produced a higher-quality DSST than the APF template. The median SNR for APF observations was $\sim$76 per pixel at the iodine wavelength region of $\sim$500 nm. To maintain only high quality data points, we removed all (7) RVs from the APF time series that had SNR < 31, equivalent to an RV error of 9 m/s. We also removed 1 APF observation that was taken within 5 minutes of 12$^{\circ}$ twilight in the morning.

\section{System Properties}
\label{systemproperties}

\subsection{Host Star}
\label{HostStar}

\begin{table*}[t]
\tabletypesize{\footnotesize}
\caption{System Parameters}
\label{tbl:totalparams}
\resizebox{\textwidth}{!}{
\begin{tabular}{lcccccc}
\hline
\hline
\multicolumn{6}{c}{\textbf{Stellar Parameters}}  \\
\hline
\textbf{Parameter} & \textbf{Value} & \textbf{Source} &  &  &  \\
\hline
\textbf{General} &  &  &  &  &  \\
Other Names &  TOI 1339, HIP 99175 & \\
RA & 20:08:06.15 & \citet{GaiaCollab}\\
Dec &  +66:51:01.08 & \citet{GaiaCollab} \\
V mag &  8.97 & \citet{BA2020} \\
\hline
\textbf{Astrometry} \\
Parallax (mas) & 18.71 $\pm$ 0.07 & \citet{BA2020} \\
Proper Motion in RA (mas) & 150.26 $\pm$ 0.04 & \citet{BA2020} \\
Proper Motion in Dec (mas) & $-63.91 \pm 0.05$ & \citet{BA2020} \\
Radial Velocity (km/s) & $-9.5 \pm 0.2$ & \citet{GaiaCollab} \\
\hline
\textbf{SpecMatch Spectroscopy} \\
$T_{\mathrm{eff}}$ (K) & 5348 $\pm$ 100 & \texttt{SpecMatch-Synthetic} \\
log \textit{g} (cm s$^{-2})$ & 4.3 $\pm$ 0.1 & \texttt{SpecMatch-Synthetic} \\
$[\mathrm{Fe/H}]$ (dex) & $-0.15 \pm 0.06$ & \texttt{SpecMatch-Synthetic} \\
$v \sin i$ (km/s) & < 2.0 & \texttt{SpecMatch-Synthetic} \\
$logR^{'}_{HK}$ (dex) & $-5.11 \pm 0.05$ & \texttt{SpecMatch-Synthetic} \\
Spectral Type & G9V & \citet{PecaultMamajek} \\
\hline
\textbf{Isochrone Modelling} \\
Radius, $R_{*} (R_{\odot})$ & 0.94 $\pm$ 0.02 & \texttt{Isoclassify} & \\
Mass, $M_{*} (M_{\odot})$ & 0.81 $\pm$ 0.04 & \texttt{Isoclassify} & Before following \citet{Tayar2020}, error was $\pm$ 0.03\\
Luminosity, $L_{*} (L_{\odot})$ & 0.65 $\pm$ 0.02 & \texttt{Isoclassify} \\
Age (Gyr) & > 8.7 & \texttt{Isoclassify} \\
\hline
\textbf{Stellar Abundances (Dex)} from \texttt{KeckSpec} \\ 
$[$C/H$]$ & $-0.12\pm0.07$ & & $[$N/H$]$ & $-0.17\pm0.09$ & \\
$[$O/H$]$ & $0.09\pm0.09$ & & $[$Na/H$]$ & $-0.18\pm0.07$ \\
$[$Mg/H$]$ & $-0.09\pm0.04$ & & $[$Al/H$]$ & $-0.02\pm0.08$ \\
$[$Si/H$]$ & $-0.11\pm0.06$ & & $[$Ca/H$]$ & $-0.17\pm0.07$ \\
$[$Ti/H$]$ & $-0.07\pm0.05$ & & $[$V/H$]$ & $-0.12\pm0.07$ \\
$[$Cr/H$]$ & $-0.26\pm0.05$ & & $[$Mn/H$]$ & $-0.38\pm0.07$ \\
$[$Ni/H$]$ & $-0.21\pm0.05$ & & $[$Y/H$]$ & $-0.17\pm0.09$ \\
\hline
\hline
\multicolumn{6}{c}{\textbf{Planet Parameters}}     \\
\hline
\textbf{Parameter} & \textbf{Planet b} & \textbf{Planet c} & \textbf{Planet d} & \textbf{Planet e} & \textbf{Planet f} \\
\hline
Orbital Period (days) & $8.88029 \pm 0.00002$ & $28.5805 \pm 0.0002$ & $38.3525 \pm 0.0003$ & $101.5 \pm 0.4$ & \DGperL-\DGperH \\
Time of Conjunction (BJD) & $2458715.3561 \pm 0.0004$ & $2458726.0534 \pm 0.0006$ & $2458743.5518 \pm 0.0007$ & $2459043.6 \pm 0.3$ & $-$ \\
Duration (hours) & $3.1 \pm 0.1$ & $4.5 \pm 0.2$ & $5.5 \pm 0.3$ & $-$ & $-$  \\
Impact Parameter & $0.62 \pm 0.02$ & $0.63 \pm 0.02$ & $0.48 \pm 0.04$ & $-$ & $-$ \\
Inclination (degrees) & $88.06 \pm 0.08$ & $89.09 \pm 0.03$ & $89.43 \pm 0.04$  & 88.0-89.4 & $-$ \\
$R_p/R_*$ & $0.0336 \pm 0.0007$ & $0.0306 \pm 0.0007$ & $0.0302 \pm 0.0007$ & $-$ & $-$ \\
Radius ($\rearth$) & $\rb$ & $\rc$ & $\rd$ & $-$ & $-$ \\
Semi-major axis (AU) & $0.078 \pm 0.001$ & $0.170 \pm 0.002$ & $0.207\pm 0.003$ & $0.397 \pm 0.005$ & \DGsemiL-\DGsemiH \\
$^*$Equilibrium temperature (K) & $893 \pm 36$ & $605 \pm 24$ & $549 \pm 22$ & $397 \pm 16$ & $-$ \\
Eccentricity & 0 (fixed) & 0 (fixed) & 0 (fixed) & 0 (fixed) & 0 (fixed) \\
RV semi-amplitude ($\ms$) & $\Kb$ & $\Kc$ & $\Kd$ & $\Ke$ & >23.0 \\
Mass ($\mearth$) & $\mb$ & $\mc$ & $<5.8$ at $2\sigma$ & $\meearth$ / $\sin i$ & 630-3500 \\
Density (g/cc) & $\rhob$ & $\rhoc$ & $\rhod$ & $-$ & $-$ \\
\hline
\hline
\multicolumn{6}{c}{\textbf{Model Parameters}}  \\
\hline
\textbf{Parameter} & \textbf{Value} & &  &  &  \\
\hline
Linear Limb Coefficient, $u_1$ & $0.42 \pm 0.06$ & & & & \\ 
Quadratic Limb Coefficient, $u_2$ & $0.05 \pm 0.08$ & & & & \\
HIRES Zeropoint, $\gamma_\mathrm{HIRES}$ ($\ms$) & $-22.26$ & & & & \\
HIRES Jitter, $\sigma_\mathrm{HIRES}$ ($\ms$) & $1.7 \pm 0.2$ & & & & \\
APF Zeropoint, $\gamma_\mathrm{APF}$ ($\ms$) & $-8.01$ & & & & \\
APF Jitter, $\sigma_\mathrm{APF}$ ($\ms$) & $3.7 \pm 0.6$ & & & & \\
Trend, $\dot{\gamma}$ ($\ms d^{-1}$) & $\gammadot$ & & & & \\
Curve, $\ddot{\gamma}$ ($\ms d^{-2}$) & $\gammaddot$ & & & & \\
\hline
\hline   
\footnotesize $^*$Equilibrium Temperatures assume zero bond albedo
\end{tabular}
}
\end{table*}

\par  We analyzed our iodine-free HIRES spectrum with the \texttt{SpecMatch-Syn} code \citep{CKS1} to derive the $T_{\mathrm{eff}}$, $\log g$, and metallicity [Fe/H] of the host star, and we list our results in Table \ref{tbl:totalparams}. We then derived stellar mass, radius, and age according to the approach described in \citet{FulPet2018}. We incorporated Gaia DR2 parallaxes \citep{GaiaCollab}, 2MASS apparent $K$ magnitude, and the MIST models \citep{Choi2016} using the \texttt{isoclassify} package \citep{Huber2017, Berger2020}. Following \citet{Tayar2020}, we inflated the error bar on the stellar mass measurement by adding a systematic error term of 0.03 $M_{\odot}$ in quadrature. Given the limited spread in the HR diagram at HD 191939’s $T_\mathrm{eff}$ (5348 $\pm$ 100 K, G9V), isochrone ages have large uncertainties. However, they indicate this star is older than 8.7 Gyr (2$\sigma$ confidence). 

\par We determined the abundances for 15 individual elements using \texttt{KeckSpec} \citep{Rice2020} finding the composition of HD 191939 is generally sub-solar for most elements. We determine the Mg/Si ratio to be consistent with both the solar value and most local stars \citep{Brewer2016}. C/O, however, is found to be $0.34 \pm 0.09$; $2\sigma$ lower than the solar value implying the assumption of solar abundances for these elements may not be applicable to stellar atmospheric models. We obtain [Y/Mg] = $-0.08 \pm 0.1$ and use this with the abundance-age relation of \citet{Nissen2020} which gives an age estimate of $7 \pm 3$ Gyr. Although this is consistent with the lower bound obtained from an isochronal fit, HD 191939 is 400K cooler than the Sun-like stars used for this relation and should be treated with caution.

\par HD 191939 is a chromospherically inactive star with $\log R^{\prime}_{HK}$ = -5.11 $\pm$ 0.05. We computed the Ca II H\&K index ($S_{HK}$) as described in \citet{IandF2010} for both our Keck/HIRES and APF time series, see Table \ref{tbl:RVData}. We find no significant correlations between the $S_{HK}$ values and RVs. Additionally, we find no statistically significant periodicities in the $S_{HK}$ time series.

\begin{figure}[t]
  \centering\includegraphics[width=0.49\textwidth]{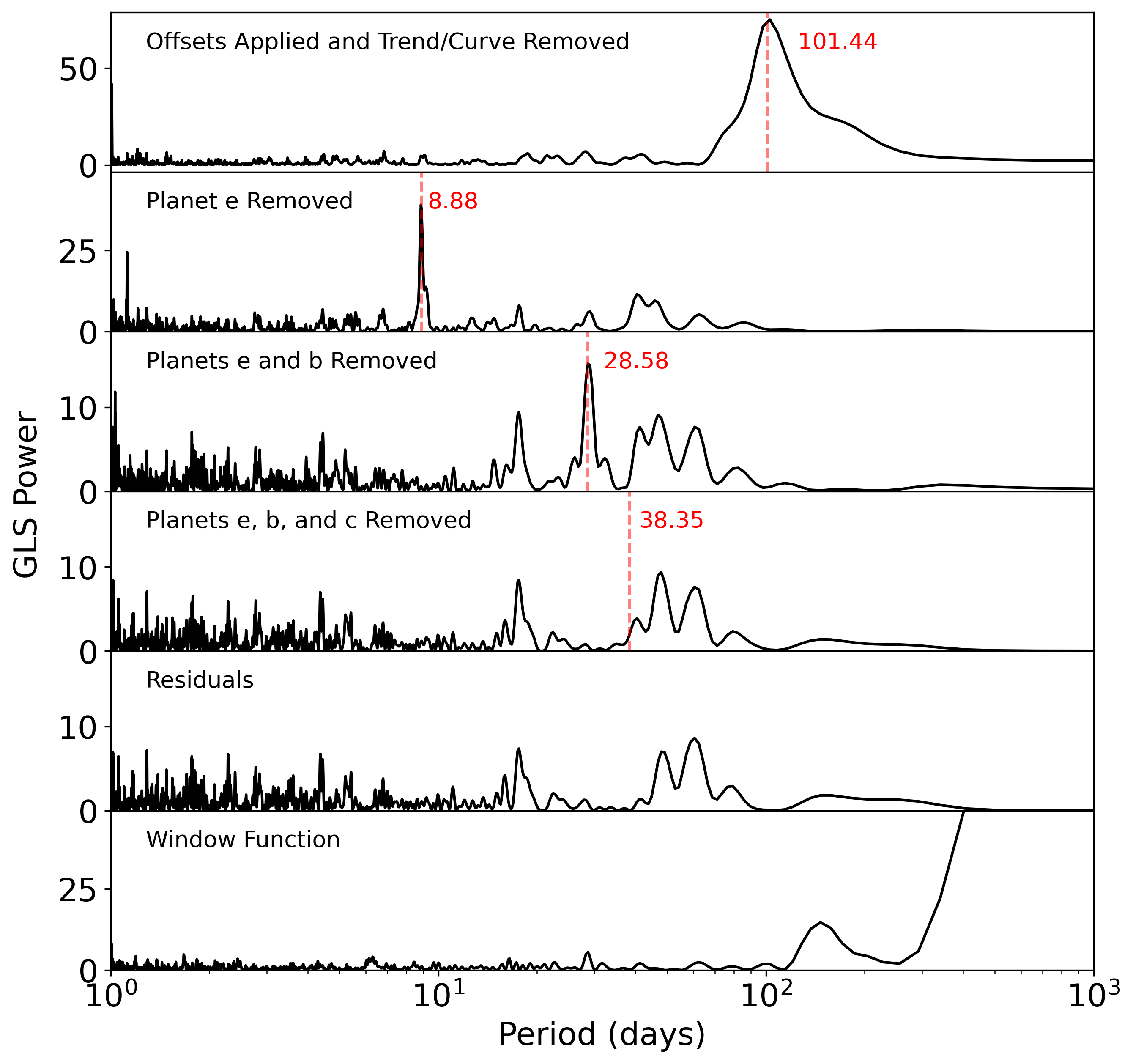}
 \centering\caption{GLS periodograms of the combined time series of Keck/HIRES and APF data. The two data sets were first filtered by removing instrumental offsets as well as the trend and curvature according to the best fit parameters from our preferred model. In each descending panel, we have removed one planet at a time. The bottom panel shows the window function of the time series.}
  \label{fig:periodograms}
\end{figure}

\begin{figure*}[t] 
  \centering\includegraphics[width=\textwidth]{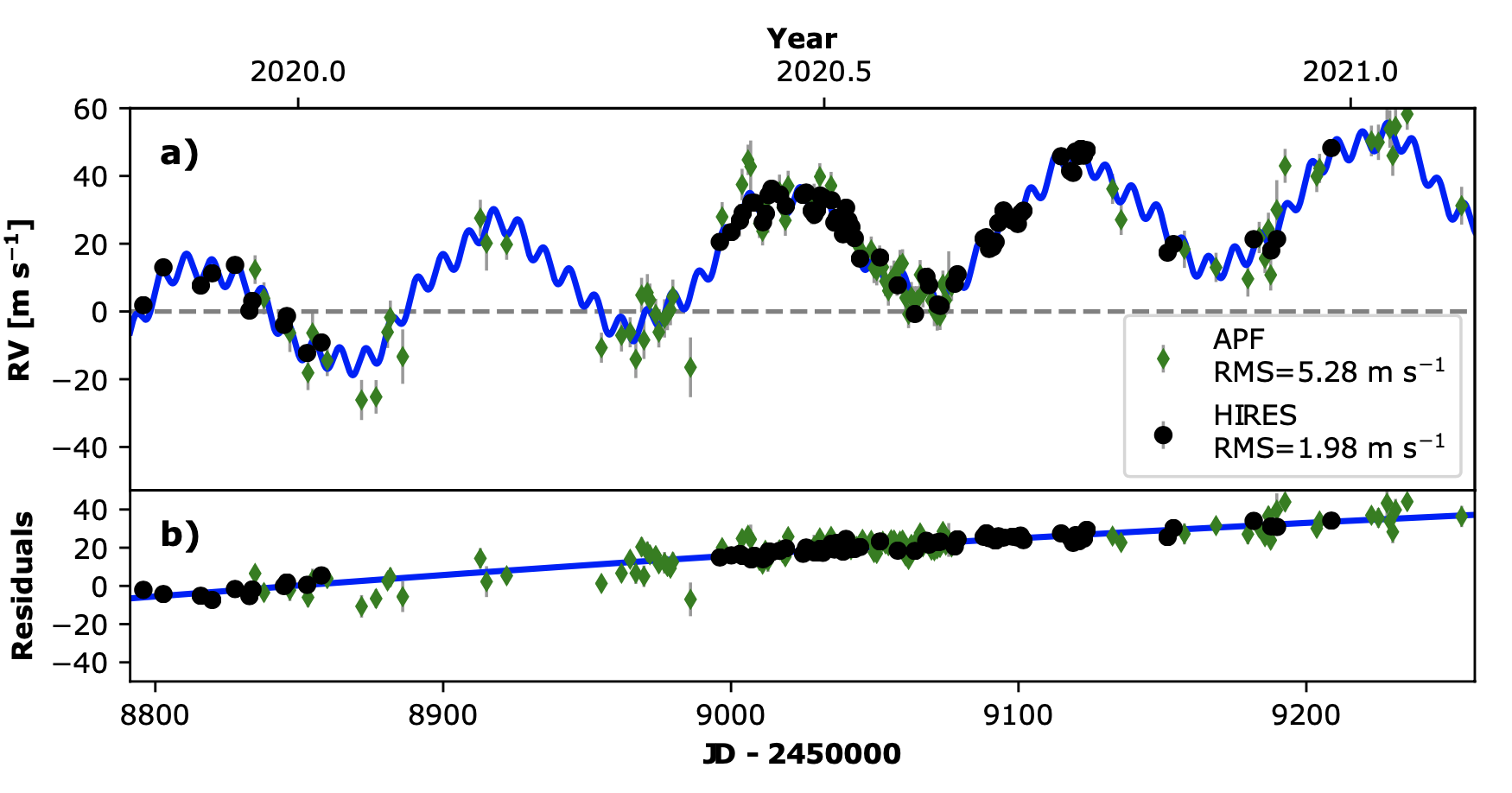}
 \centering\caption{\textbf{a)} Our complete RV time series with our preferred model (blue) as well as \textbf{b)} residuals including trend and curvature. Data collected from Keck/HIRES are shown as black circles while data from the APF are shown by green diamonds.}
  \label{fig:timeseries}
\end{figure*}

\begin{figure*}[b]
 \centering\includegraphics[width=0.7\textwidth]{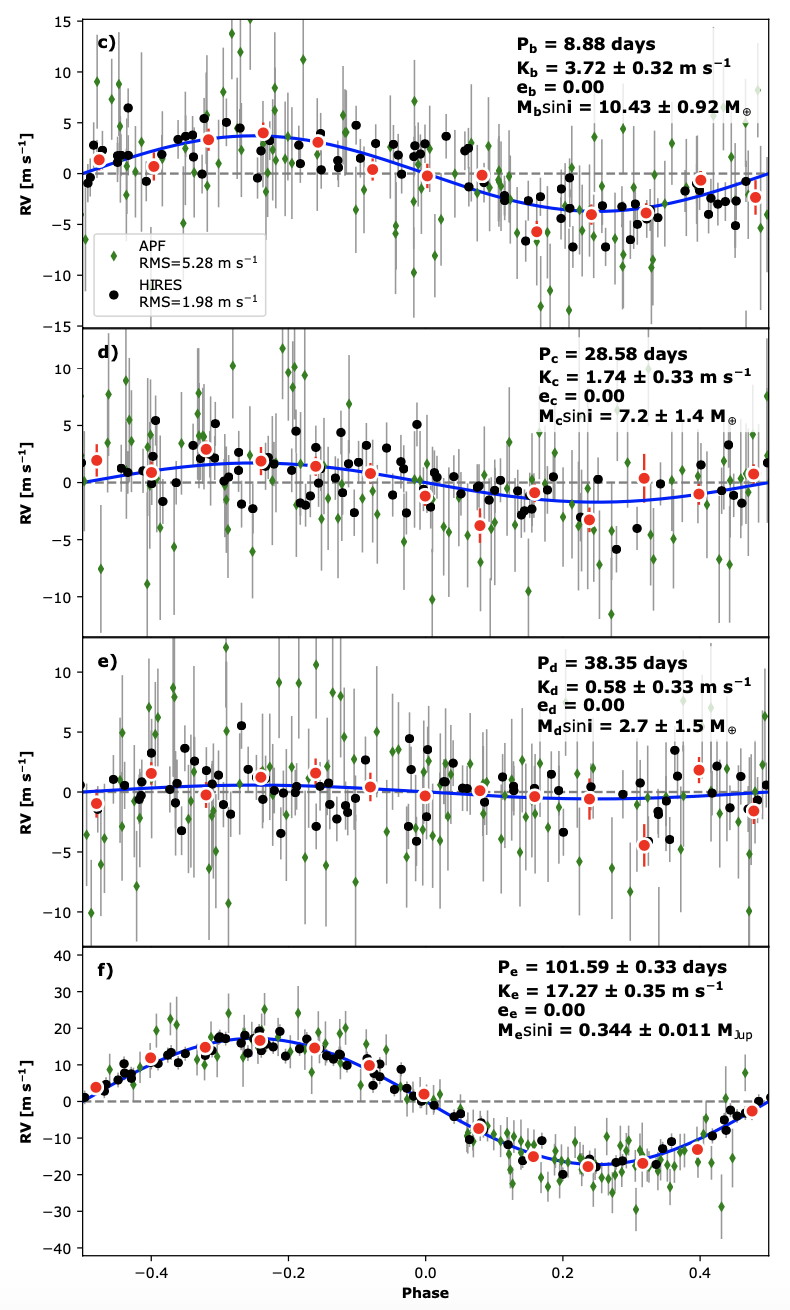}
 \centering\caption{The phase folded RV time series for each planet with periods less than our baseline. Red circles are bins of size 0.08 phase.}
  \label{fig:RVzoomins}
\end{figure*}

\subsection{RV Model}
\label{ModelComp}

\par Soon after beginning our RV observations, we saw evidence of an additional planet beyond those identified by $\tess$, including observations which showed a $\sim40$ m/s change in the RV, consistent with a massive planet. Continued observations further traced a large amplitude periodicity near $\sim100$ days. The Generalized Lomb-Scargle (GLS) periodogram \citep{ZechmeisterKurster} of the RVs is dominated by this signal which we attribute to a fourth planet (e) (Figure \ref{fig:periodograms}, top panel).

\par To discern the architecture of the system, we performed a model comparison analysis. Using \texttt{RadVel} \citep{radvel}, we tested a variety of RV models: 3-5 total planets and either allowing eccentricity to vary for each or fixing it to zero, as well as allowing or prohibiting trend and/or curvature terms. We used Markov Chain Monte Carlo (MCMC) to explore the parameter space and estimate uncertainties; all planet models discussed here converged by the default \texttt{RadVel} criteria unless otherwise stated. 
\par In all models, we fixed the periods and times of conjunction of the transiting planets to the values found from our $\tess$ photometry model. We set uniform priors on the Doppler amplitudes ($-\infty$, $\infty$), allowing negative values for all planets to avoid biasing the masses to higher values. We set a uniform prior from 1 to 1000 days on the period of planet e and a uniform prior on its time of conjunction (2459000, 2459100) BJD. For both instruments, we set a prior on the instrumental jitter as uniform (0, 10) m/s. Lastly we set a prior on the trend term uniform from (-1, 1) m/s/d and on the curvature term uniform (-0.1, 0.1) m/s/d$^2$.

Our preferred RV model has an Akaike Information Criterion (AIC) \citep{Akaike1974} of 858 and contains four planets on circular orbits, as well as both a trend and curvature term, which models a subset of a sinusoid as a quadratic to represent a 5th body in the system. The closest neighboring model, in terms of AIC, is one with 4 planets plus a trend but no curvature term (AIC = 866). Our preferred model is very strongly preferred over a model with 3 transiting planets (did not converge, AIC = 1332) and a 4-planet model with no trend and no curvature (AIC = 1202). Our full RV timeseries can be seen in Figure \ref{fig:timeseries}, and phase-folded RV time series for each planet can be seen in Figure \ref{fig:RVzoomins}. The orbital parameters and masses of all planets can be found in Table \ref{tbl:totalparams}. We searched the residuals of our preferred model for additional planets but found no statistically significant signals.

\begin{figure*}[t]
  \centering\includegraphics[width=\textwidth]{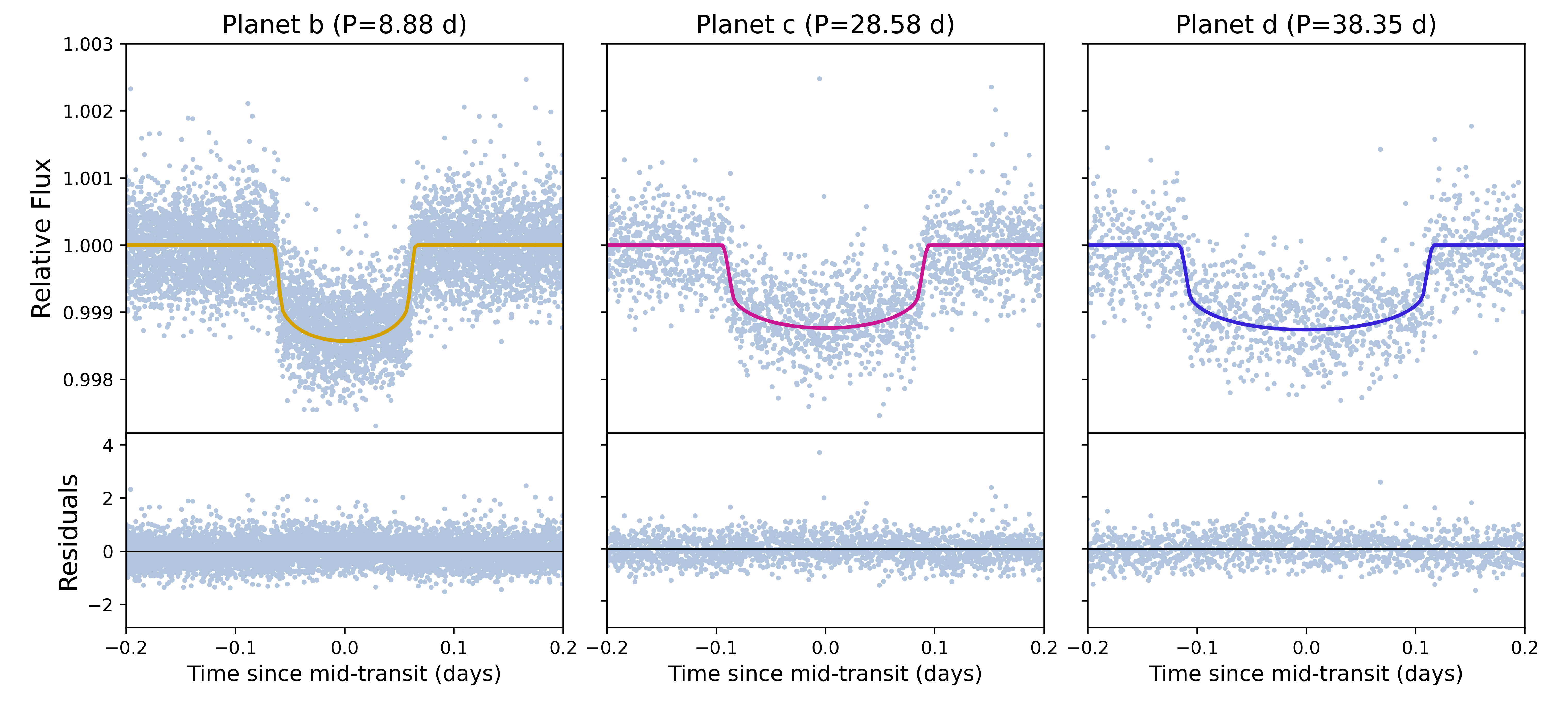}
 \centering\caption{Phase folded light curves for each of the transiting planets with our best fit model overlaid and residuals below.}
  \label{fig:phasefold}
\end{figure*}

\begin{figure*}[h!]
  \centering\includegraphics[width=0.7\textwidth]{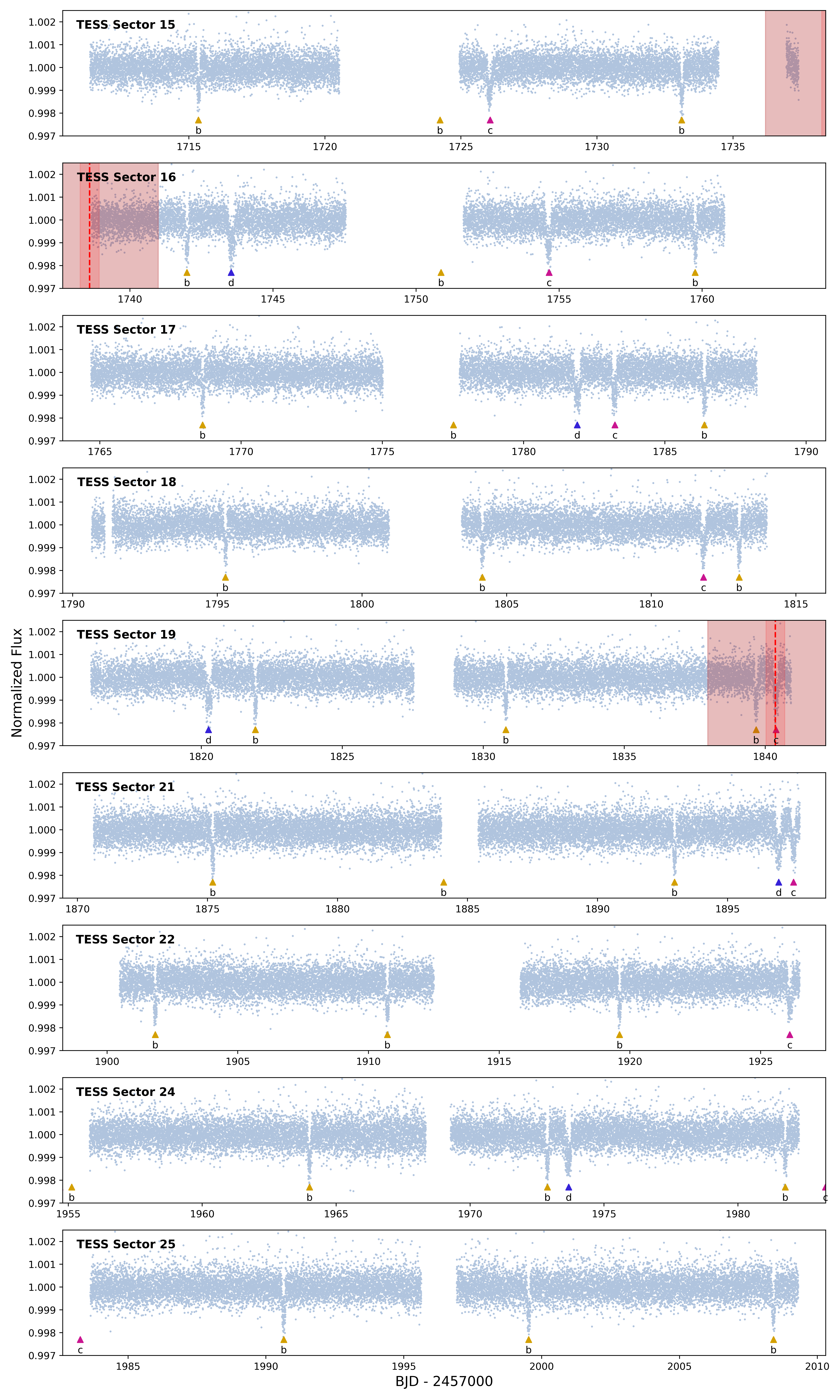}
 \centering\caption{$\tess$ photometry from sectors 15-19, 21, 22, 24, and 25 highlighting the transits of the three sub-Neptunes which are indicated by color-coded arrows. Our RV model's predicted transit midpoint times for planet e are shown by vertical dashed red lines along with $3\sigma$ error windows as light red shaded regions. The predicted $\sim8$ hour transit duration (for a central transit) is shown by dark red shading. An additional transit window occurred in Sector 23 when the star was not visible in any $\tess$ cameras.}
  \label{fig:allphotometry}
\end{figure*}

\subsection{Photometry Model}
\label{photometrymodel}

\par We pre-whitened the $\tess$ photometry by using a Gaussian process model to subtract out low amplitude stellar and instrumental variability from the light curve. We then performed a blind transit search using \texttt{Transit Least Squares} (\texttt{TLS}; \citealt{TLS}). This recovered three transiting planets with period, depth, and duration values and errors consistent with the previously published values in \cite{BA2020}. We also performed a more targeted search with \texttt{TLS} for transits of planet e but found no evidence of any such events. We then modeled the transits of planets b, c, and d with the \texttt{Exoplanet} package \citep{exoplanetDFM} and re-derived planet parameters using our updated stellar parameters and the \texttt{TLS} output as priors (Table \ref{tbl:totalparams}).

\par To calculate this model, we assumed circular orbits and fit for seventeen parameters: (1.) Orbital periods, with Gaussian priors informed by our \texttt{TLS} search values, (2.) Times of inferior conjunction, with Gaussian priors informed by our \texttt{TLS} search, (3.) Planet-to-star radius ratios, with a log-uniform prior from 0.01 to 0.1, (4.) Impact parameters, with a uniform prior from 0 to 1, (5.) Stellar radius, with Gaussian priors defined by the updated stellar parameters, (6.) Stellar mass, with Gaussian priors defined by the updated stellar parameters, (7.) Quadratic limb darkening parameters calculated using \texttt{Python Limb Darkening Toolkit} \citep{ldtk2015}, and (8.) A white noise scaling term for the $\tess$ light curve. \texttt{Exoplanet} implements MCMC algorithm, which we ran with 2000 iterations and a 500 step burn in and found that all chains converged. Additionally, we derived transit midpoints for each transit of each planet which is discussed further in \S\ref{mmr} (see Appendix, Table \ref{tbl:TTVs}). Figure \ref{fig:phasefold} shows our modeled phase-folded light curves for each planet and Figure \ref{fig:allphotometry} shows the full reduced $\tess$ light curve of the star with transits color coded.

\par Given our weak detection of planet d in the RV time series, we returned to the photometry to confirm the period and transit times. Due to the positioning of data gaps in the light curve, there are 4 "odd" transits and 1 "even" transit of planet d. We considered the possibility that the single even transit, occurring at 2458781.89 BJD, comes from a different source than the four odd transits. In such a scenario, the orbital period of planet d would double to 76.7 days. However, comparing between the transits, including matching depths, durations, and ingress/egress shapes, we found no inconsistencies between the single transit and the other four. Furthermore, we find no evidence for a $\sim$76 day periodicity in our RV time series. Thus we conclude that all five transits do originate from a single planet with an orbital period of 38.4 days.

\begin{figure}[h!]
    \includegraphics[width=0.47\textwidth]{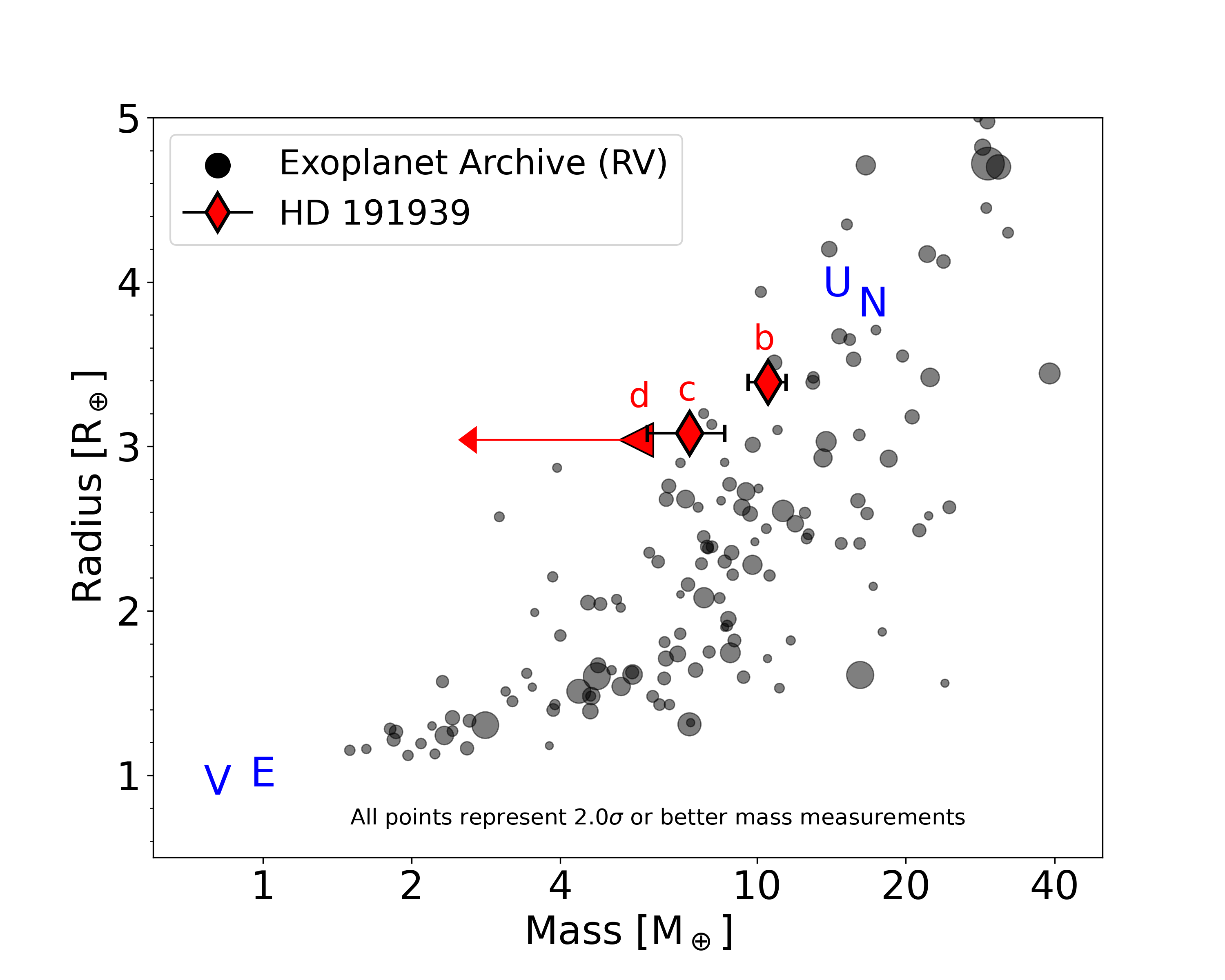}
 \centering\caption{A Mass-Radius diagram highlighting the HD 191939 transiting planets. Larger marker sizes correspond to more precise mass measurements, excluding the HD 191939 planets. Planet d's marker represents the 2$\sigma$ upper limit, and its arrow points back to the median value. Grey points are from the NASA Exoplanet Archive as of 2021-07-01, with cuts to include only 2$\sigma$ masses or better.}
  \label{fig:massradius}
\end{figure}

\section{Composition of Transiting Planets}
\label{bulkcomposition}

\par How do the transiting planets in this system compare to other known transiting planets? We find planet b imparts a Doppler semi-amplitude of $\Kb$ ms$^{-1}$, corresponding to a mass of $\mb \mearth$; plant c imparts $\Kc$ ms$^{-1}$, corresponding to $\mc \mearth$; and planet d imparts $\Kd$ ms$^{-1}$, corresponding to $\md \mearth$. The placement of the three transiting planets on a mass-radius diagram reveals that they exist at the periphery of the known planet population (Figure \ref{fig:massradius}). Planet b fits more consistently with previously known planets, while planet d is a low-mass outlier. The relatively low masses for their radii implies small densities. We find planet b has a bulk density of $\rhob$ g/cc, planet c has $\rhoc$ g/cc, and planet d has $\rhod$ g/cc. 

\par \citet{Fulton2017} and \citet{VanEylen2018} described the radius gap as a region of radius phase space from 1.5-2.0 $\rearth$ where relatively few planets are found. Studies have explained this gap as most likely due to a transitional phase between planets with and without extended H/He envelopes, which may be due to photoevaporation \citep{Lopez2014, OwenWu2017}. Given that all three transiting planets in the HD 191939 system have radii above the gap, it is likely that the best description of their compositions is that of a volatile rich envelope surrounding a rocky core \citep{Weiss2014, Rogers2015, Fulton2017}. Employing \texttt{Smint} \citep{smint}, which interpolates the model grids from \citet{Lopez2014} and \citet{Zeng2016} and samples posterior space with MCMC, we explored the possible fractions of H/He by mass for the three transiting planets assuming a dry, Earth-like, rock-iron core. Using a flat prior for the age from 9 to 13 Gyr, we find H/He envelopes of \atmb\% for planet b, \atmc\% for planet c, and \atmd\% for planet d. 

\par From our RV model, we place a 2$\sigma$ upper limit on planet d's mass at 5.8 \mearth. This corresponds to 2$\sigma$ upper limit on planet d's density of 1.1 g/cc. While this density upper limit places it within the range of planets b and c, the potential low density for planet d is noteworthy. In the literature, there is a population of low density planets: the Kepler-51 system \citep{Masuda2014}, Kepler-79d \citep{JontofHutter2014}, and Kepler-87c \citep{Ofir2014}, which are collectively described as "super-puffs" for their inflated radii (4-8 $\rearth$) and low masses (2-5 $\mearth$), which implies densities of $\sim$0.1 g/cc. While HD 191939 d is not a super-puff since its radius is smaller (only 3 \rearth), it does share a notable characteristic with the super-puffs: they all exist in or near resonance with another planet in their systems. The super-puff planets may have low masses for their sizes as part of a selection bias: the planet masses are derived from transit timing variation (TTV) interactions, which are most prominent for planets in or near a resonance chain. HD 191939 d's potential low density, combined with its placement as the outer member of a near 4:3 resonance with planet c (see \S\ref{mmr} for more detail), draws some comparison to the super-puffs and brings forward questions on its possible formation history.

\par Two different mechanisms have been proposed for explaining the prevalence of highly inflated plants in or near resonance. \citet{LeeChiang2016} showed super-puff planets most easily gain their extended atmospheres in dust-free environments at distances beyond 1 AU before migrating inwards. As part of this migration, they are more likely to form the outer companion of a resonance chain with another interior planet in the system. Under this formation scenario, planet d would likely contain a large fraction of water, a composition which we do not explore in this paper. \citet{Millholland2019} describes how super-puffs that exist just wide of resonance with another planet are thought to have preferentially high obliquities, which could drive heat dissipation through obliquity tides resulting in inflated planet radii.

\par  HD 191939 d represents a unique opportunity to study a possible low density planet and to test the above theories for two reasons. The mass measurement we provide comes from the RV method rather than TTVs. The location in the system interior to the Jovian planet e can provide dynamical constraints for any potential migration  history. Of the super-puffs listed above, only Kepler-79d has a confirmed planet exterior to its orbit in the system, and this planet is another sub-Neptune.

\par The relatively small masses, low densities, and high equilibrium temperatures of these planets might combine to drive atmospheric escape on some or all of the three inner planets. By the Jeans escape mechanism, to first order approximation a gas will eventually completely escape if its thermal velocity exceeds one sixth the planet's escape velocity. Planet b's temperature is likely high enough to allow the steady escape of atomic and molecular hydrogen. Fixing each planet's radius to the median values of our photometry model, we calculated whether molecular hydrogen would escape each planet for a grid of every combination of planet mass and equilibrium temperature out to 3$\sigma$ of each value. We find that molecular hydrogen escapes planet b in 84\% of combinations, 52\% for planet c, and 94\% for planet d. Following the same procedure, planet d's small mass means it may not even be able to retain helium as 47\% of combinations allow this gas to escape. If any of these planets are experiencing atmospheric escape, transmission spectroscopy with $\jw$ might show evidence. 

\section{Planet f Constraints}
\label{5thcomp}

\par What is the nature of the 5th planet in the system? Our RV analysis favors both a trend and curvature in the residuals of the preferred 4-planet model, suggesting a 5th planet with an orbital period much longer than our 415-day observing baseline. The presence of this planet can be further constrained by the change in HD191939's proper motion over a period of 24 years. Using these independent data sets, we can place constraints on the mass and semi-major axis of planet f.

\par We derived these constraints using a novel method which compares model orbits using just 3 free parameters. We quantify long-period signals in the RV residuals through trend ($\dot{\gamma}$) and curvature ($\ddot{\gamma}$) terms; and astrometric motion through $\Delta \mu$, the difference in proper motions at two epochs. We generated a set of randomly-sampled orbits and computed these three parameters for each. A high-likelihood orbital model is one that reproduces the true values of $\dot{\gamma}$, $\ddot{\gamma}$, and $\Delta \mu$.

\par To produce a set of model orbits, we first defined our search range for both mass and semi-major axis. We started with $\tau_{min}$, the lower bound on orbital period. Planet f produced only a small detected curvature over our observing baseline, a feature that we estimate would require an orbital period $\gtrsim$ 4 times the baseline. This yielded a lower semi-major axis limit of 2.6 AU. We limited our search to semi-major axes within 50 AU. We used a similar argument to obtain a lower bound on $M_p$. We took the maximum $\Delta$RV from the residuals of fitting for planets b-e and set it equal to the semi-amplitude of a planet with a period of $\tau_{min}$, again assuming a circular orbit. From this amplitude, we calculated a minimum mass of 2.05 $\mj$. We chose 200 $\mj$ as the upper limit of our mass search, reasoning that more massive objects would be luminous enough to detect in high-contrast imaging.

\par We marginalized over four additional orbital parameters: inclination $i$, eccentricity $e$, argument of periastron $\omega$, and mean anomaly $M$. In total we drew $10^8$ random samples from this 6-dimensional parameter space using the following prior distributions:

\begin{itemize}
    \item{$\log\left ( \frac{a}{\text{1 AU}} \right ) \sim
    \mathcal{U}(2.62, 50)$}
    \item{$\log\left ( \frac{M_p}{\text{$1 \mj$}} \right ) \sim
    \mathcal{U}(2.05, 200)$}
    \item{$\cos(i) \sim \mathcal{U}(0, 1)$}
    \item{$\omega \sim \mathcal{U}(0, 2\pi)$}
    \item{$M \sim \mathcal{U}(0, 2\pi)$}
    \item{$e \sim \mathcal{B}(0.867, 3.03)$}
\end{itemize}

\noindent where $\mathcal{B}$ is the two-parameter \cite{kipping2013} beta distribution for $e$. We used the same samples to generate both the RV curves and the astrometric proper motions.

\par To impose RV constraints, we computed for each sample the first ($\dot{\gamma}$) and second ($\ddot{\gamma}$) time derivatives of the stellar radial velocity. We began by differentiating the true anomaly $\nu$:

\begin{gather}
    \nu = 2 \tan^{-1} \left ( \sqrt{\frac{1+e}{1-e}} \tan \left ( \frac{E}{2} \right )\right )
    \label{eq:nu}
\end{gather}

\begin{gather}
    \dot{\nu} = \frac{2 \pi \sqrt{1-e^2}}{\tau \left ( 1-e\cos(E) \right )^2},
    \label{eq:nu_dot}
\end{gather}

\noindent where $\tau$ is the orbital period calculated from Kepler's Third Law and E is the eccentric anomaly, which we obtained by numerically solving Kepler's equation:

\begin{gather}
    M = E - e\sin E.
    \label{eq:kepler}
\end{gather}

\noindent The second derivative of $\nu$ is also needed to compute $\ddot{\gamma}$:

\begin{gather}
    \ddot{\nu} = -\dot{\nu}^2 \frac{2e\sin(E)}{\sqrt{1-e^2}}
    \label{eq:nu_ddot}
\end{gather}

With the derivatives of $\nu$, we can write the equations for $\dot{\gamma}$ and $\ddot{\gamma}$. We start with the RV value itself, $\gamma$:

\begin{gather}
    \gamma = K \left [ e\cos(\omega) + \cos(\nu + \omega) \right ],
    \label{eq:gamma}
\end{gather}

\noindent where 

\begin{gather}
    K = \sqrt{\frac{G}{1-e^2}}\frac{M_p \sin i}{\sqrt{a(M_p+M_\star)}}.
    \label{eq:K}
\end{gather}

The derivatives of $\gamma$ are:

\begin{gather}
    \dot{\gamma} = -K \left [ \dot{\nu} \sin(\nu + \omega) \right ]
    \label{eq:gamma_dot}
\end{gather}

and

\begin{gather}
    \ddot{\gamma} = -K \left [ \dot{\nu}^2 \cos(\nu + \omega) + \ddot{\nu} \sin(\nu + \omega) \right ].
    \label{eq:gamma_ddot}
\end{gather}

We evaluated the sample likelihood according to

\begin{gather}
    P(\dot{\gamma}, \ddot{\gamma}|\dot{\gamma}_m, \ddot{\gamma}_m) \propto
    \text{exp}\left[-\left(\frac{(\dot{\gamma} - \dot{\gamma}_m)^2}
                         {2\sigma_{\dot{\gamma}}^2}
                  + \frac{ (\ddot{\gamma} - \ddot{\gamma}_m)^2}
                         {2\sigma_{\ddot{\gamma}}^2}
                         \right)\right].
    \label{eq:likelihood_RV}
\end{gather}

\begin{figure}[t]
  \centering\includegraphics[width=0.45\textwidth]{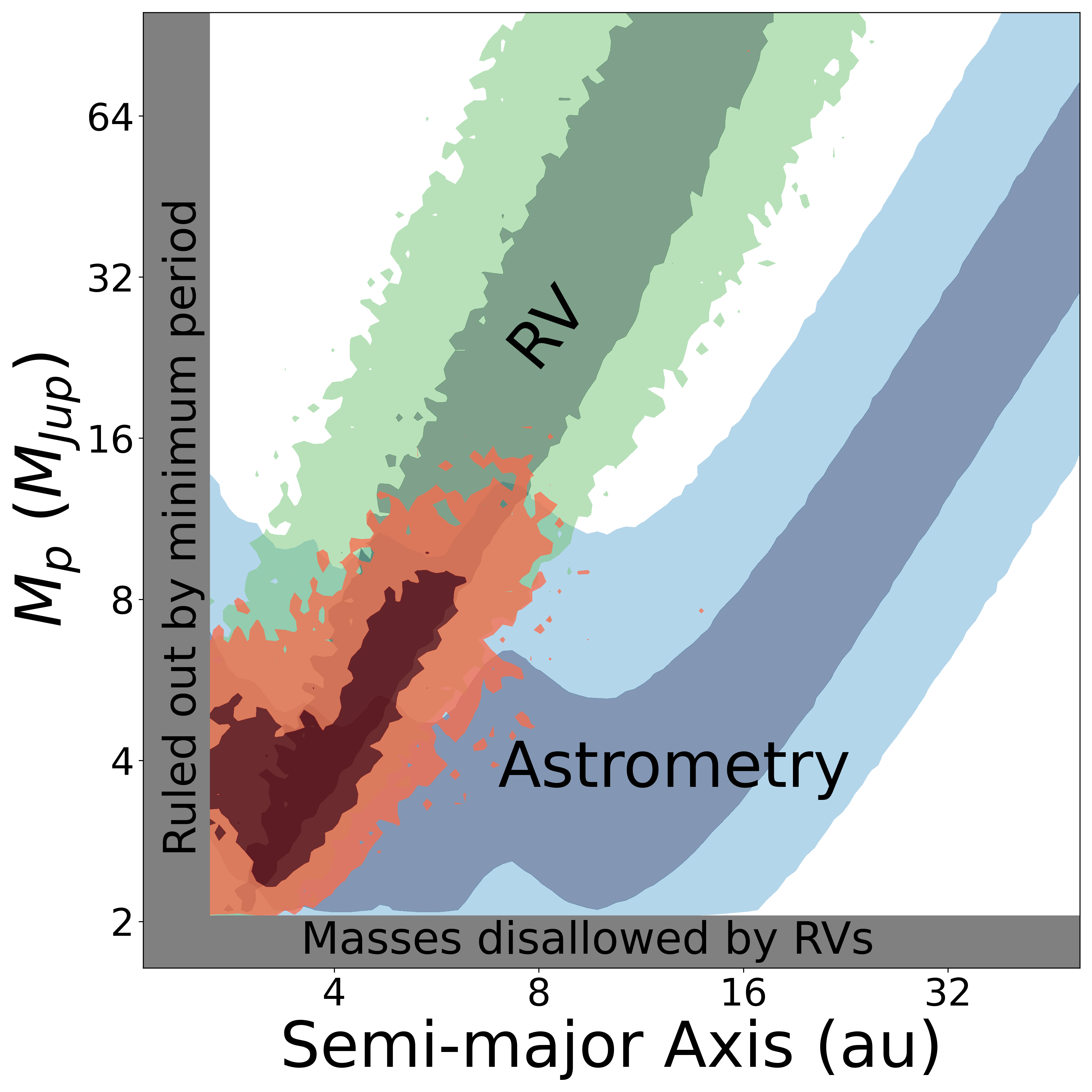}
 \centering\caption{Constraints on the mass and semi-major axis of planet f. The green region shows values that are consistent with the measured RV trend and curvature. The blue region shows values that are consistent with the Hipparcos/Gaia astrometry. The red region shows the values consistent with both RV and astrometry. Dark and light regions indicate the 1 and 2$\sigma$ confidence intervals, respectively. Planet f is likely between \DGmassL--\DGmassH \, $M_J$, orbiting between \DGsemiL--\DGsemiH \, AU.}
  \label{fig:5thcompanion}
\end{figure}

To obtain the 2D $a$-$M_p$ joint  posterior, we marginalized over $\{e, i, \omega, M\}$. The results from the RV only constraints can be seen in Figure \ref{fig:5thcompanion} in green with 1- and 2-$\sigma$ contours.

\par We next incorporated astrometry to further constrain the characteristics of the fifth planet. \citet{Brandt2021} aligned the reference frames of \textit{Hipparcos} \citep{HipparcosCatalog} and \textit{Gaia EDR3} \citep{GaiaEDR3} to produce a self-consistent catalog of stellar proper motions measured at epochs 1991.25 and 2015.5. \citeauthor{Brandt2021} reported the proper motion based on the difference in position between these epochs. The \textit{Gaia} and position-derived proper motions, $\vec{\mu}_G$ = ($150.19 \pm 0.02, -63.99 \pm 0.02$) mas/yr and $\vec{\mu}_{HG}$ = ($150.31 \pm 0.03, -63.94 \pm 0.03$) mas/yr, were the most precise, and indicated a change in proper motion $\Delta \mu = |\vec{\mu}_G - \vec{\mu}_{HG}|$ of $0.13 \pm 0.03$ mas/yr over the 24 years separating the two epochs.

\par Using the same orbit models as in the RV analysis, we first computed the average proper motion vector in the \textit{Gaia EDR3} epoch. We also used the change in astrometric position between the \textit{Gaia} and \textit{Hipparcos} epochs to obtain an average proper motion over the 24 year baseline. We then computed the magnitude of the difference vector $\Delta \mu_m$ and evaluated the likelihood via

\begin{gather}
    P(\Delta \mu|\Delta \mu_m) \propto
    \text{exp}\left(-\frac{(\Delta \mu - \Delta \mu_m)^2}{2\sigma_{\Delta \mu}^2}\right).
    \label{eq:likelihood_astro}
\end{gather}

\par The detected proper motion difference rules out high mass models that were permitted by our RV-only analysis. The blue region of Figure \ref{fig:5thcompanion} shows the range of \textit{a}-\textit{$M_p$} values that are allowed by astrometry at the 1 and 2$\sigma$ levels. 

\par Because the RV and astrometric data sets are independent, we may evaluate the joint RV-astrometry likelihood by multiplying Equations \ref{eq:likelihood_RV} and \ref{eq:likelihood_astro}. Figure \ref{fig:5thcompanion} shows in red the region of $a$-$M_p$ space that is allowed by both the RV and astrometric constraints. We find at $95\%$ confidence that planet f has a mass of \DGmassL--\DGmassH \, $\mj$ and orbits at a distance of \DGsemiL--\DGsemiH \, AU.

Throughout this paper we refer to this companion as a ``planet" because these current mass constraints place it most likely below the generally accepted upper mass limit for planets of $\sim13 \mj$; but we caution that the high-mass tail of the probability distribution includes objects that would typically be characterized as brown dwarfs. Such high mass objects on the planet-brown dwarf boundary are thought to form by one of two general formation pathways: core accretion \citep{Pollack96} or gravitational instability \citep{Boss97}. Core accretion is more successful at producing low mass objects and is the most plausible formation channel for planets b through e. \citet{Schlaufman2018} showed a transition point in formation mechanism at $10 \mj$, which may represent a mass upper limit for objects formed via core-accretion. Therefore, more massive objects more likely formed via gravitational instability and are therefore not planets. If planet f is at the upper end of its mass range, gravitational instability becomes a plausible pathway. This raises the possibility that both mechanisms were active in the HD 191939 system. We advocate for continued Doppler/astrometric monitoring of the HD 191939 system to fully resolve this companion's orbit and measure its mass more precisely to identify which formation channel is more likely.

\section{Planet e is Nearly Coplanar}
\label{LLsection}

\par What is the inclination of planet e? Given the emergence of planet e in our RV data, we searched the $\tess$ photometry for evidence of its transit. We would expect this $\me \mj / \sin i$ Jovian planet to have a radius of $\sim1 \rj$, implying a transit depth on the order of 1\%. At a 101 day orbital period, assuming zero eccentricity and an edge-on orbit, we expect the duration of its transit to be $\sim8$ hours. Such a transit event should be obvious in the data by visual inspection. We do not see planet e's transit (see Figure \ref{fig:allphotometry}).

\par Within the error bars of our period and time of conjunction for planet e, it is possible that $\tess$ missed the transits of planet e by unlucky timing. Still, the most likely explanation for the missing transits is that the planet is non-transiting. We did not search for a transit of planet f because its transit event should be a similar depth but even longer than planet e's and it was not near its expected time of conjunction at the time of $\tess$'s observations.

\par Assuming planet e is non-transiting and has a radius of 1 $R_J$, we place an upper limit on the inclination at  $89.5^{\circ}$. To place a lower limit, we explored the dynamics of the system with Laplace-Lagrange secular perturbation theory \citep{LaplaceOriginal}. Following the methods in \citet{murray_dermott_2010}, we analytically derived equations for the time dependence of the inclination for each of the planets in the system. We chose to ignore effects from planet f. Due to planet f's large semi-major axis relative to the other 4 planets, the inner 4 will move together under its influence. Additionally, any of effects from planet f will play out over much longer timescales than we are interested in ($\sim$2  orders of magnitude longer). For the four planets in question, we used the median values for mass and semi-major axis from Table \ref{tbl:totalparams}. Within the Laplace-Lagrange framework, eccentricity and inclination become decoupled; for simplicity and consistency with our preferred RV model, we assumed circular orbits.

\begin{figure*}[t]
  \centering\includegraphics[width=0.47\textwidth]{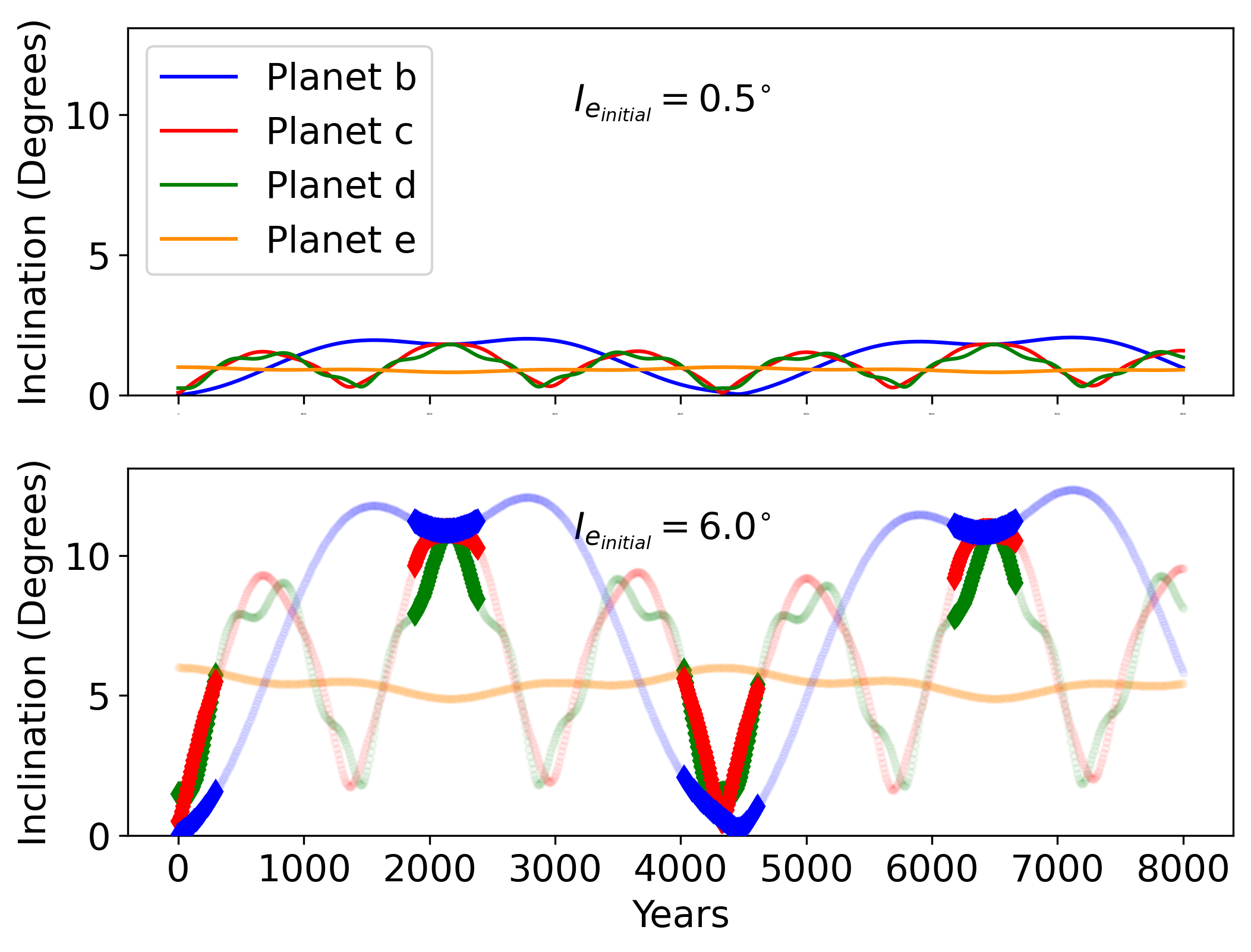}
   \centering\includegraphics[width=0.45\textwidth]{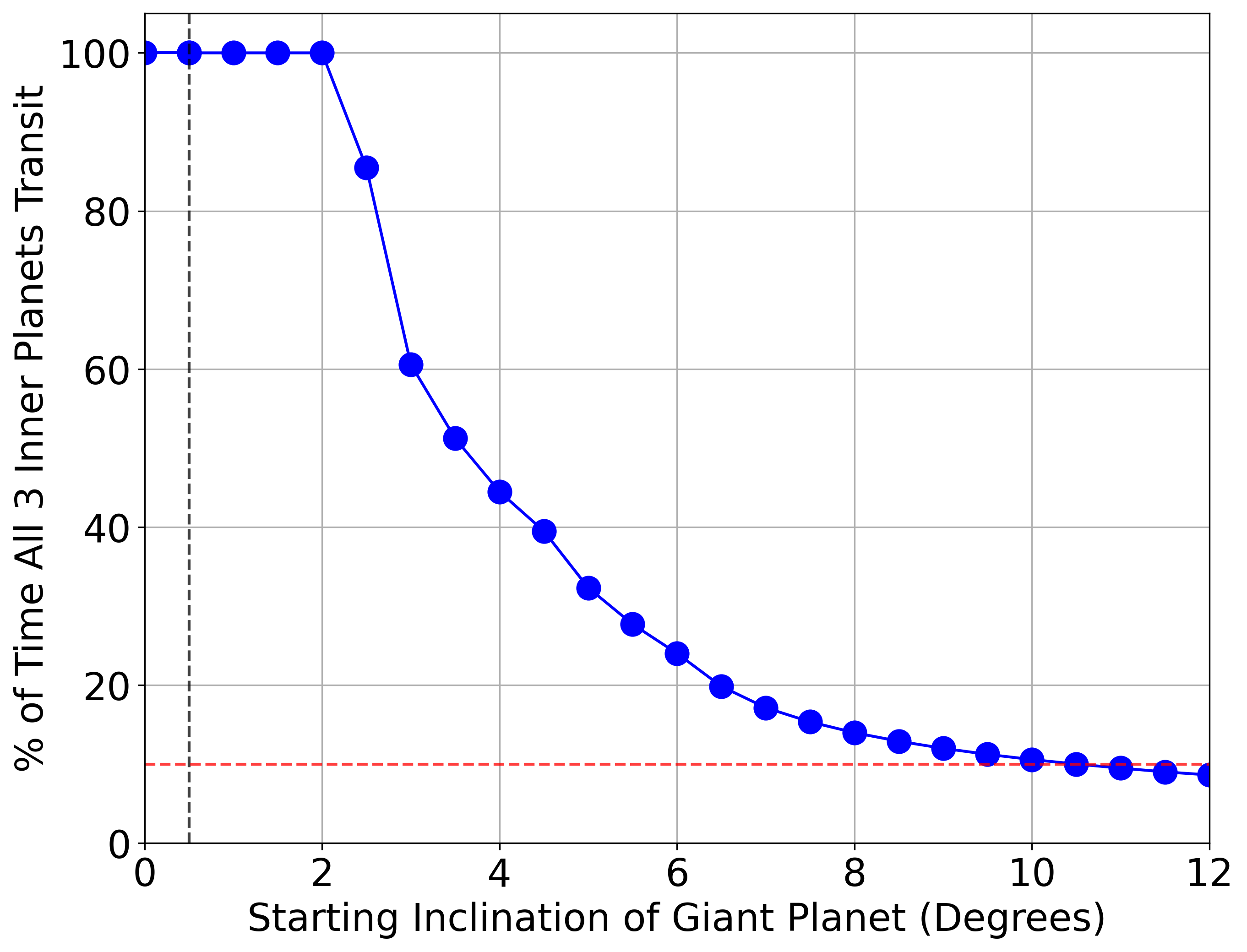}
 \centering\caption{Details on the Laplace-Lagrange analysis. \textbf{Left:} The inclination curves for each planet when planet e is given a starting value of $I_e = 0.5^{\circ}$ vs $I_e = 6.0^{\circ}$. When the mutual inclination of the three is small enough for all three to transit together, the line is opaque. \textbf{Right:} The percent of time during which the inner three planets transit depends on the inclination of planet e. The vertical black dashed line indicates the nominal maximum inclination for which we would expect planet e to still transit. The horizontal red dashed indicates the 10\% threshold for our conservative estimate on the upper limit to the giant planet's inclination. }
  \label{fig:LLdetails}
\end{figure*}

\par The Laplace-Lagrange secular perturbation theory is built on the foundation of the disturbing function, where \textit{I} is the inclination, \textit{j} and \textit{k} are planet indices that run from 0 to N with N being the number of planets in the system: 

\begin{gather}
    \frac{\partial I_j}{\partial t} = -\frac{1}{n_j a^2_j I_j} \frac{\partial R_j}{\partial \Omega_j},
    \label{eq:disturb}
\end{gather}

\noindent where $R_j$ is the disturbing function

\begin{gather}
    R_j = n_j a^2_j \left [ \frac{1}{2} B_{jj}I^2_j +B_{jk}I_j I_k \cos(\Omega_j - \Omega_k)  \right ]
    \label{eq:disturb2}
\end{gather} 

\noindent and 

\begin{gather}
    B_{jk} = \frac{1}{4}\left [  \frac{G(M_* + m_j)}{a^3_j} \right ] ^\frac{1}{2}   \frac{m_k}{M_* + m_j}  \alpha_{jk}  \overline{\alpha}_{jk}   b^{(1)}_{\frac{3}{2}}(\alpha_{jk}),
    \label{eq:LL3}
\end{gather} 

\noindent and

\begin{gather}
   n_j = \sqrt{\frac{G(M_* + m_j)}{a_j^3}}
\end{gather}

\noindent where $B_{jk} = -B_{jj}$. Terms $\alpha_{jk}$ and  $\overline{\alpha}_{jk}$ are constants determined by semi-major axis ratios of the $j$th and $k$th planets, $b^{(1)}_{\frac{3}{2}}(\alpha_{jk})$ is a definite integral also dependent on semi-major axes \citep{murray_dermott_2010}, and $\Omega$ is the longitude of ascending node. From the disturbing function we constructed the B matrix. The eigenvalues of the B matrix, $f_k$, represent the periodicity of the oscillations of the planets' inclination and the eigenvectors (which are unscaled and must be normalized) along with the initial conditions of the system's configuration represent the amplitude of the oscillations. 

\par In the normalization process we calculated both a scaling factor and a phase angle for the oscillation periodicity of each planet, $\gamma_k$. This is accomplished by implementing the initial conditions at t = 0 (both $I_o$ and $\Omega_o$) to generate a set of set of \textit{2N} equations from which we can solve for N scaling factors and N phase angles. With these scaling factors in hand, the final amplitudes of the oscillations, $V_{jk}$, are determined.

\par Then we calculated the inclinations of each planet at a given time: \\ 

\begin{gather}
    I_j = (p^2_j + q^2_j)^{\frac{1}{2}} ,
    \label{eq:LL4}
\end{gather} 

\noindent where $p_j$ and $q_j$ are parameterized variables:

\begin{gather}
    p_j = \sum_{k=0}^{N_{planets}} V_{jk} \sin(f_k t + \gamma_k),
    \label{eq:LL5}
\end{gather} 

\begin{gather}
    q_j = \sum_{k=0}^{N_{planets}} V_{jk} \cos(f_k t + \gamma_k).
    \label{eq:LL6}
\end{gather}

\par Within this framework, we derived $I_j(t)$ for each planets j $\in$ \{b, c, d, e\} for various initial configurations of the system.

\par For each configuration, planets b, c, and d were initialized at $0^{\circ}$, corresponding to placing all three on the same plane. Note that the plane from which we are measuring inclinations is $90^{\circ}$ transposed from the conventional plane of reference for inclinations, the sky plane. For ease of reference, we call this plane the \textit{LL-Plane}. We also initialized all four planets' longitude of ascending node, $\Omega$, to the same value, arbitrarily $0^{\circ}$. We tested various trials where $\Omega_e$ was initialized at different values between $0^{\circ}$-$360^{\circ}$ and found it had little to no affect on the outcome of our experiment. In each configuration we set the starting inclination for planet e to different values, stepping in $0.5^{\circ}$ intervals from $0^{\circ}$ to $12.0^{\circ}$.

\par We computed $I_j(t)$ for an 8,000 year span, roughly double the longest eigenfrequency. For every year in a configuration, we computed the mutual inclination of the three planets:

\begin{gather}
    \cos I_{xy} = \cos I_x \cos I_y + \sin I_x \sin I_y \cos(\Omega_x - \Omega_y),
    \label{eq:LL7}
\end{gather} 

\noindent \citep{Carter2012}. We determined a maximum limiting angle for mutual transiting of the inner 3 planets by geometric reasoning. We calculated the minimum transiting inclinations for both the innermost and second innermost planets, by $i_{min} \approx \frac{R_*}{a}$. Then the sum of these two angles is the limiting angle. This corresponds to placing the innermost and second innermost planets at the opposite limbs of the star. For a given timestamp, if the mutual inclinations of all pairs of planets are less than the limiting angle, then all planets transit together at that timestamp.

\par Figure \ref{fig:LLdetails} shows the $I_j(t)$ curves for two examples from our trials as well as the results of all trials. For each trial of planet e's starting inclination, we computed the percent of timestamps within the 8000 year time span during which all three of the inner planets transited with respect to an arbitrary line of sight. As expected, the farther from the LL-Plane that we start planet e's inclination, the smaller the percent of the timestamps during which all three inner planets will transit. There is a range of starting inclinations for which we would expect all three inner planets to transit 100\% of the timestamps, from $0^{\circ}$ to $2.0^{\circ}$ in the LL-Plane. We nominally rule out inclinations less than $0.5^{\circ}$ based on the absence of a transit for planet e, although this limit does not take into account the uncertainty in planet e's radius and the simplification that all three inner planets start at $0^{\circ}$. In sample tests where we included planet f with mass and semi-major axis values drawn from results in \S\ref{5thcomp}, we find the results to be similar. Including planet f, the value for the percentage of timestamps where the inner 3 planets are all transiting for any given inclination of planet e is within 5\% of the value as when we exclude planet f.

\par Above $2.0^{\circ}$ in the LL-plane, the percentage of timestamps where all three are transiting together falls sharply and then decreases asymptotically towards 0\%. From these results, we conservatively place a upper limit on the planet e's mutual inclination at $10^{\circ}$. This angle corresponds to a lower limit for absolute inclination of $80^{\circ}$ in the conventional sky-plane frame of reference. For starting inclinations above $10^{\circ}$, the amplitudes of the planets' oscillations in inclination space become large enough that it is rare for all three to transit together from an arbitrary line of sight: $<10\%$ of the timestamps tested. Mutual inclinations of planet e larger than $10^{\circ}$ are viable solutions. However, in those scenarios, the decreasingly short windows in time where all three planets transit make Earth observers increasingly lucky to have caught the system at one of these rare moments in its dynamical periodicity. This investigation suggests that planet e is likely to be nearly coplanar with the three transiting planets.

\section{TTVs and MMR}
\label{mmr}

\par Planets c and d have orbital periods very near to 4:3 mean-motion resonance (MMR). But do they indeed reside in MMR? We explored this possibility and the implications which follow.

\par In general, planets which reside in MMR are characterized by period ratios of 

\begin{gather}
    \frac{P_{2}}{P_{1}} = \frac{j}{j-1},
    \label{ttv1}
\end{gather}

\noindent where $j$ is an integer and subscripts 1 and 2 denote the inner and outer planet of the pair, respectively. We quantify the "proximity" to MMR by 

\begin{gather}
    \Delta_{12} = \frac{P_{2}}{P_{1}}\frac{j-1}{j} - 1, 
    \label{ttv2}
\end{gather} 

\noindent following \cite{Lithwick2012}.

Applying this formula to planets c and d, $\Delta_{cd} = 0.6432 \pm 0.0001\%$. Following \citet{Batygin2017}, the resonant bandwidth can be approximated as:

\begin{gather}
    \left | \chi  \right | \underset\sim < 5 \frac{j-1}{j^{2/3}} \Big(\frac{M_1 + M_2}{M_*}\Big)^{2/3}.
    \label{ttv3}
\end{gather}

For planets c and d, $\chi_{cd} = 0.662 \pm 0.001\%$. Because $\Delta < \left | \chi  \right |$, we cannot rule out that the two planets are librating in MMR. 

\par Under the assumption that planets c and d are close to but \textit{not} in MMR, we calculated the period and amplitude of TTV oscillations of the pair following \citet{Lithwick2012}. TTV oscillations will be oppositely-phased sinusoids, each at a period designated as the \textit{super period }(SP): 

\begin{gather}
    P_{SP} = \frac{P_{2}}{j \left | \Delta \right |},
    \label{ttv4}
\end{gather} 

\noindent with amplitudes 

\begin{gather}
    TTV_{1} = P_{1}\frac{ ( \frac{m_{2}}{M_{*}} )}{\pi j^{2/3}(j-1)^{1/3}\Delta}\left(-f -\frac{3Z}{2\Delta}\right) ,
    \label{ttv5}
\end{gather} 

\noindent and 

\begin{gather}
    TTV_{2} = P_{2}\frac{ ( \frac{m_{1}}{M_{*}} )}{\pi j\Delta}\left(-g -\frac{3Z}{2\Delta}\right),
    \label{ttv6}
\end{gather}

\noindent where \textit{f} and \textit{g} are constants associated with the MMR ratio, in this case 4:3, and Z is a linear combination of the free eccentricities of the two planets.

\par We calculated the super period of planets c and d to be 1490 $\pm$ 10 days. In the circular orbit limit, Z = 0 and the amplitudes of planet c and d's TTV oscillations are $15.5 \pm 9.1$ minutes and $59.2 \pm 13.8$ minutes, respectively. If the phase of the oscillations is near an inflection point, Planet d's oscillation would be large enough that it could be detected even though $\tess$ has only sampled about a fifth of the super period.  

\par To further investigate, we calculated the TTV associated with each transit event. We generated model transits offset from the expected transit time by between $\pm$60 minutes and calculated the chi-squared ($\chi^2$) fit of these model transits to the light curve. We adopted the offset that minimized the $\chi^2$ statistic as the value of the TTV. The 1$\sigma$ error bars are calculated from the offset where the $\chi^2$ increased from its minimum value by 1.0. We performed this process for each transit of each planet. 

\par
\begin{figure}[t]
  \centering\includegraphics[width=0.45\textwidth]{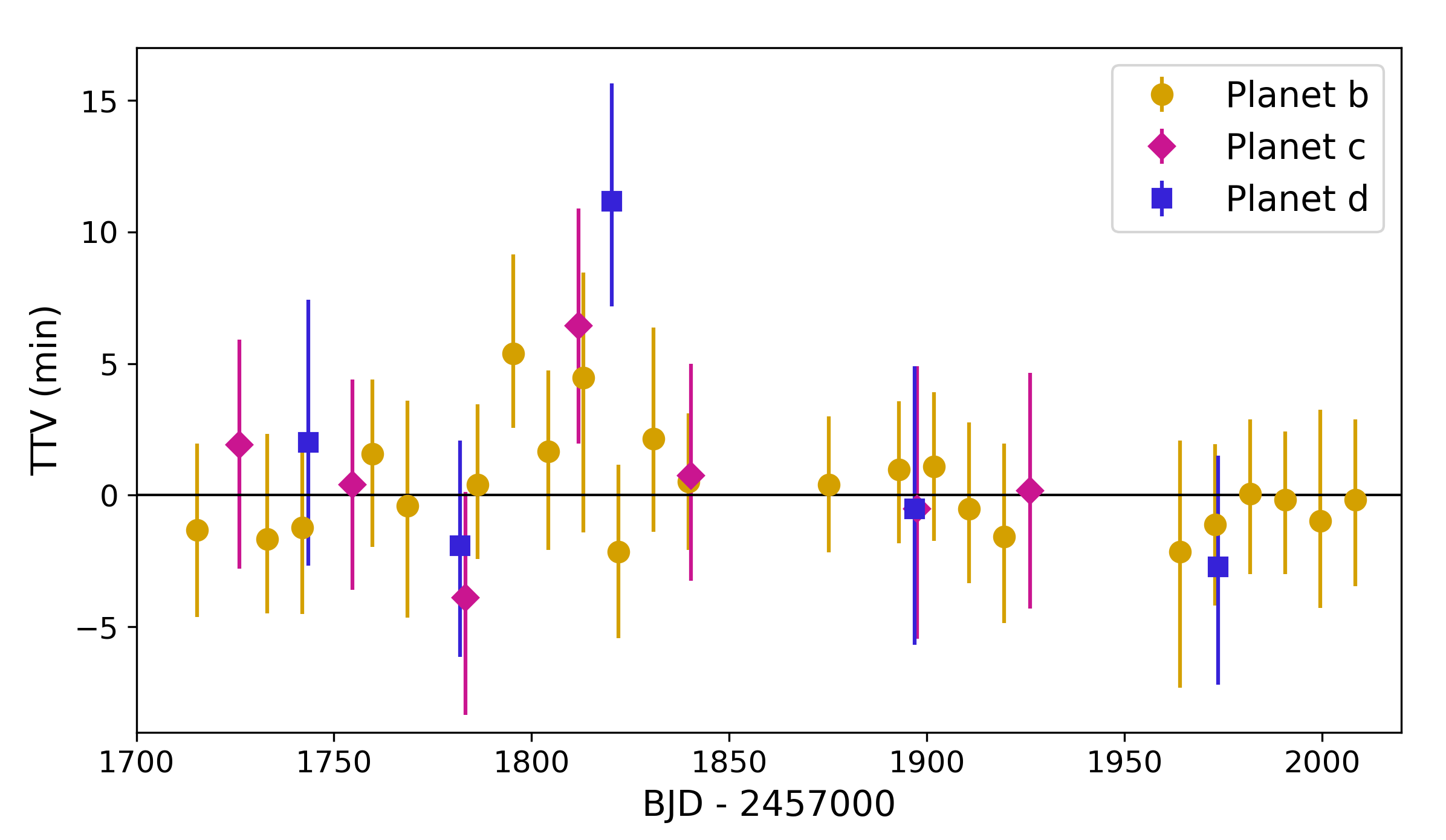}
 \centering\caption{TTVs of the transiting planets over the duration of the $\tess$ photometry. We do not detect significant TTVs for any of the transiting planets over the observing baseline.}
  \label{fig:allttvs}
\end{figure}

 Figure \ref{fig:allttvs} as well as Table \ref{tbl:TTVs} shows all of the TTVs for each planet. Planet d's 5 transits cover $\sim$230 days of time, or about 15\% of the super period. Its TTVs do not show a trend. Planet c's transits similarly span only $\sim$230d. Due to $\tess$'s observing strategy, planet c transited just hours before sector 24 observations and hours before and after sector 25 observations, at times when the star was not visible to $\tess$. It is noteworthy that the two planets behave similarly in that when one is late, the corresponding transit of the other is similarly late and vice versa for early transits. Planet b's TTVs are consistent with zero, showing no trend or significant sinusoidal variation. 

\par These results can be interpreted in two ways. First, and most likely, $\tess$ has not sampled enough of the 1500 day super period to make a conclusive finding. Alternatively, we could be sampling TTVs very near the maximum or minimum of the TTV signal's phase, so the $\Delta$TTV over the baseline is too small for a significant detection. $\tess$'s extended mission cycle 4 will shed more light onto these three possibilities. 

\section{Gap Complexity}
\label{gapcomplex}

\par Could there be an additional planet hiding in the gap between planets b and c? With planets c and d very near MMR, it is noticeable that there are not more pairs of planets also spaced in near resonant orbits. Following the peas-in-a-pod architecture where multi-planet systems show similarly sized planets in regular orbital distance spacing, we might expect more than just one pair in this system to exhibit near-resonance, especially considering that the transiting planets have very similar radii \citep{Leleu2021}.

\par In the residuals of our GLS periodogram (Figure \ref{fig:periodograms}), there is a noticeable peak between planets b and c at 17.7 days. A planet at this period would be particularly interesting as it would be near 2:1 resonance with planet b and 8:5 resonance with planet c. A planet at this period would also fill the gap in log Period space of this system well. Given that we have a strong RV detection of planet c, any additional planet in this gap between planets b and c must be less massive than planet c and inclined. When we add a fit for a 17.7d planet in our preferred model, we find a 2$\sigma$ upper limit to its mass to be 6 $\mearth$. In order to be non-transiting, its inclination must be at least 2$^{\circ}$ from the LL plane. 

\par We followed the methods in \citet{GilbertandFabrycky2020} to calculate the Gap Complexity, \textit{$\mathcal{C}$}, for the HD 191939 system. $\mathcal{C}$ describes the deviation from uniform planet spacing in a system. $\mathcal{C}$ = 0 indicates uniform spacing in log Period space, while as $\mathcal{C} \rightarrow 1$ the less uniform the spacing. For Kepler systems, $\mathcal{C}$ peaks at 0 with the majority ($\sim$75\%) of systems having $\mathcal{C} < 0.2$. Systems with larger $\mathcal{C}$ values are more likely to have additional planets hiding in the gaps between known planets. We calculate $\mathcal{C}_{HD191939} = 0.846$ considering the transiting planets only, as planet e does not fall into the peas-in-a-pod configuration. We interpret the high value of $\mathcal{C}$ to mean that there is a significant gap, which could be the site of an additional planet. When we include a hypothetical planet on a 17.7 day period with the known transiting planets, we calculate $\mathcal{C}_{HD 191939}$ = 0.18. This value is consistent with the findings of \citet{GilbertandFabrycky2020} for the general pattern of multi-planet system configurations. Adding a 17.7 day planet to our preferred model does not improve the likelihood enough to justify the extra three parameters. Nevertheless, this planet candidate is interesting and deserves continued attention with additional RV observations.

\begin{figure}[t]
  \centering\includegraphics[width=0.45\textwidth]{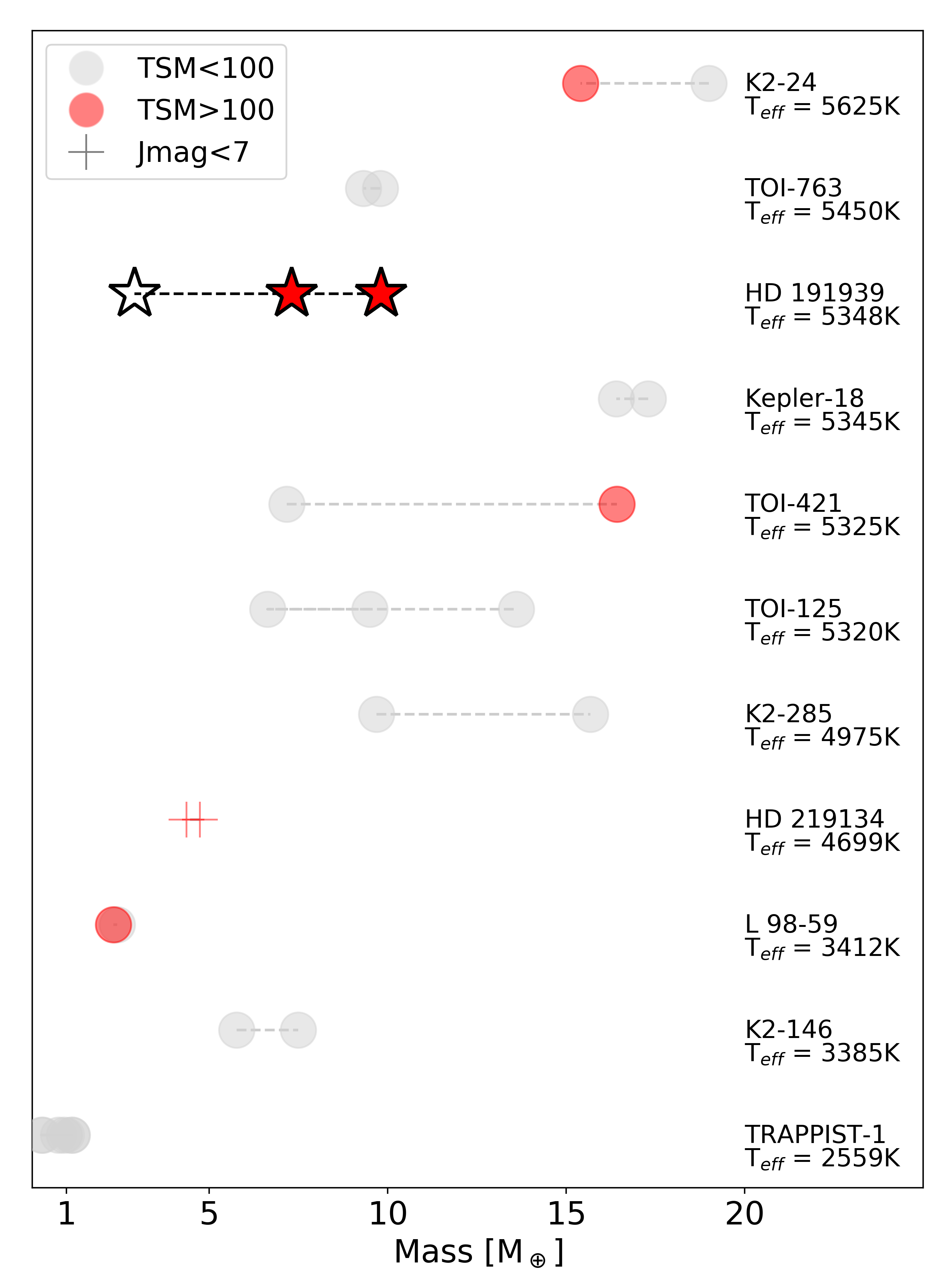}
 \centering\caption{All multi-planet systems with 5$\sigma$ masses and radii for small planets ($R_p < 10 \rearth, M_p < 100 \mearth$) with TSMs > 20. Planets are plotted by mass and arranged vertically in order of host star effective temperature (hotter at the top). HD 191939 b and c have TSM values that are individually among the best in the sub-Neptune population, and are unique in having the same host star. Due to Planet d's weak mass measurement, it appears in this plot unfilled. HD 191939 is the only system to date with multiple planets with TSMs greater than 100 that also does not saturate $\jw$.}.
  \label{fig:tsmprospects}
\end{figure}

\section{Follow Up Prospects}
\label{AtmosphericProspects}

\par How well suited is this system for further follow up? We identified HD 191939 as a key TKS target for atmospheric follow up with the target selection algorithm described in Scarsdale et al. (in prep). As a bright ($J = 7.6$ mag) multi-planet system, space-based spectroscopic observations offer a unique opportunity for studies in planet formation and evolution.

\par We use the Transmission Spectroscopy Metric (TSM; \citealt{Kempton2018}) to quantify the expected signal-to-noise ratio of $\jw$-NIRISS observations for the transiting planets:
\begin{equation} \label{eqn:tsm}
\centering
\mathrm{TSM}_\mathrm{p} = S \times \frac{R_p^3   T_{eq}}{M_p  R_*^2} \times 10^{-0.2m_J},
\end{equation}
where $S$ is a dimensionless normalization constant, equal to 1.28 for planets $2.75 < R_p < 4.0$ \rearth.
The TSM is a proxy for the expected SNR from a 10-hour observing program with $\jw$-NIRISS assuming a cloud-free, solar-metallicity, H$_2$-dominated atmosphere. For reference, HD 3167 c, a sub-Neptune orbiting an early-K dwarf with a recent water vapor detection from five \emph{HST}-WFC3 transits \citep{mikal-evans21}, has a TSM of about 100.

\par Using the derived planet parameters from Table \ref{tbl:totalparams}, we find HD 191939 b has a TSM of \TSMb, which places it in the top quartile of targets in the $2.75 < R_\mathrm{p} < 4.0$ \rearth\ range from the statistical sample in \cite{Kempton2018}. HD 191939 c has a TSM of \TSMc, placing it in the third quartile from the top of TSM values for planets between $2.75$ and $4.0$ \rearth. We place a lower limit on the TSM of planet d, finding TSM$_\mathrm{d}$ \TSMd at 2-$\sigma$ confidence. For the transit durations reported in Table \ref{tbl:totalparams}, our TSM values scale to an expected single-transit SNR with \jw-NIRISS of \SNROneTransitb, \SNROneTransitc, and \SNROneTransitd for planets b, c, and d respectively, where the lower limit for planet d represents 2-$\sigma$ confidence.

\par We used \texttt{PandExo} \citep{batalha17} to estimate the nominal heights of molecular features in a single-transit \emph{JWST}-NIRISS transmission spectrum for planet b, assuming a cloud-free, solar-metallicity atmosphere. In this ideal case we find feature heights of $\sim$100-300 ppm between 1 and 5 $\mu$m. In reality, clouds and/or enhanced atmospheric metallicity will probably reduce these amplitudes by a factor of three or more \citep{Wakeford19}. Additionally a sub-Solar C/O ratio, which may be implied from the host star's abundance measurements, also disagrees with the ideal case of a solar-metallicity composition and would produce spectra dominated by CO, H$_2$O, and CO$_2$.

\par A spin-orbit measurement for this system would be particularly informative to planetary formation theories. Only 8 systems with three or more planets have had their sky-projected obliquity angles, $\lambda$, measured. In the HD 191939 system, the three inner planets all lie in nearly the same orbital plane, while we have shown that the giant planet should lie close to this plane. If they are misaligned with respect to the stellar spin axis, that could inform the dynamical history of the system and the roles that planets e and f have played in shaping the system. However, the low $v \sin i$ (see Table \ref{tbl:totalparams}) of the host star might be prohibitive to a Rossiter-McLaughlin (RM; \citealt{Rossiter1924, McLaughlin1924, GaudiWinn2007}) measurement of even the largest expected signal from planet b. A simulation using \texttt{arome} \citep{Boue2013} finds that for $v \sin i$ = 1 km/s and $\lambda = 0^{\circ}$, planet b's expected RM amplitude is 1.5 m/s. 

\par HD 191939 will be observed again by $\tess$ in Cycle 4. Nominal dates for observations include 6 sectors of additional coverage: 41, 48, 49, 51, 52, and 55. These observations will extend the total baseline of photometry observations to 2022-09-01 for a a total of 1142 days, about 76\% of the super period between planets c and d. 

\section{Conclusions}
\label{conclusion}

\par The overall architecture of the HD 191939 system $-$ multiple small planets, then a warm Saturn, followed by a high mass planet $-$ seemingly stands alone among known systems. Sub-Neptunes are near ubiquitous \citep{Howard2012, Petigura2013}, but the \textit{a priori} occurrence rate for warm sub-Jovians ($30-300 \, \mearth$ at 0.1$-$1.0 AU) is much smaller at $\sim$3\%, and similarly at $\sim$5\% for cold super-Jovians ($300-6000 \, \mearth$ at 3$-$10 AU) \citep{Fulton2021}. We cannot simply multiply together these occurrence rates to discern how rare it is for such a system like HD 191939 to exist, as \cite{Weiss2018} found that adjacent planets tend to have similar sizes, and some studies have found a relationship between sub-Neptune occurrence and giant planet occurrence \citep{Zhu2018, Bryan2019}

\par We searched the literature for analog systems by performing cuts on the known population for systems with 4 planets, with three sub-Neptunes ($M_p < 25 \mearth$) interior to a warm Saturn ($50 \mearth < M_p < 300 \mearth$, with orbital period of 50$-$360 days) and a long period high mass planet. However, there are a few systems that stand out as notable.

\par \cite{Mills2019} describe three systems, Kepler-65, Kepler-68, and Kepler-25 with high mass outer planets. Kepler-65 has a tight inner system of three sub-Neptunes and a 0.28 $\mj$ planet with an orbital period of 258 days, similar to the inner system of HD 191939, but there is no evidence for a trend over a $\sim$2000 day baseline. Kepler-25 is similar in having two inner sub-Neptunes in/near resonance (2:1) and a Saturn mass planet at just over a 100 day orbit; but again, no evidence for a long period companion represented by trend over its $\sim$3000 day observing baseline. Kepler-68 may represent the most similar system to HD 191939. It has an inner system of of two sub-Neptunes, then a Jovian with an orbital period of 634 days, and then strong evidence for curvature in the residuals. \citet{Mills2019} attribute this curvature to an object with a period much longer than the $\sim$3000 day baseline and place a lower limit of 0.6 $\mj$, but no upper limit. Lastly, Kepler-129 \citep{Zhang2021} bears resemblance to HD 191939 in having two inner planets at $< 45 \, \mearth$ and a high mass Jovian (8.3 \mj) on $\sim$7 year orbit. \citet{Zhang2021} also discusses the perturbations of inclinations of the inner transiting planets due to the long period Jovian. Each of these systems has pieces of the HD 191939 system, but none have the full architecture.

\par Bright, multi-planet systems are invaluable to the exoplanet community due to their enhanced follow up opportunities and comparative planet prospects. With photometry from $\tess$ and RV data from both Keck/HIRES and the APF, we have characterized the HD 191939 system: 3 transiting sub-Neptune planets, a fourth Jovian, and 5th high mass planet. We have measured the planets' masses, as well as their radii and densities where applicable. Because of our strong mass measurements of 3 of the 4 inner planets (>5$\sigma$), we are able to explore and further investigate many aspects of the system to answer more detailed questions about the system. Our main conclusions are as follows:

\begin{itemize}

    \item The bulk densities of the transiting planets are $\rho_b = $ $\rhob$ g/cc, $\rho_c = $ $\rhoc$ g/cc, and $\rho_d = $ $\rhod$ g/cc. We find the compositions of the planets are best explained by extended H/He atmospheres. 

    \item By new technique for constraining the mass and period of distant companions using both RV and astrometric data sets, we find planet f to be between \DGmassL--\DGmassH \, $\mj$ on a \DGperL--\DGperH \, day orbital period at 95\% confidence.
    
    \item Through a dynamical analysis using Laplace-Lagrange secular perturbation theory, we constrain the inclination of the non-transiting planet e. We find it most likely orbits within a plane less than $10^{\circ}$ from the plane roughly shared by the three transiting planets.
    
    \item By investigation into the potential mean motion resonance of planets c and d, we predict their TTV amplitudes to be $15.5 \pm 9.1$ minutes and $59.2 \pm 13.8$ minutes, respectively over a super period of 1490 $\pm$ 10 days. However, we find no evidence for significant TTVs over the short observing baseline (326 days) compared to the super period of the interaction (1500 days).
    
    \item We analyze of the RV residuals and Gap Complexity of the system to investigate the potential for additional planets in the system, identifying a possible planet candidate at 17.7 days which deserves continued attention.
    
    \item We evaluate the transiting planets' prospects for atmospheric characterization through transmission spectroscopy with $\jw$. HD 191939 is the only system that does not saturate JWST-NIRISS where two planets both have TSMs greater than 100, making it an excellent candidate for comparative atmospheric studies.
    
\end{itemize}

\par With its three transiting mini-Neptunes, one non-transiting Jovian planet, and distant high mass planet surrounding a bright, nearby host star, HD 191939 provides a rich natural laboratory for detailed atmospheric characterization and dynamical studies. \\

\textit{Facilities}
Automated Planet Finder (Levy), Keck I (HIRES), TESS \\

\textit{Software:}
\texttt{Astropy} \citep{astropy2013},
\texttt{corner.py} \citep{dfm2016}, 
\texttt{emcee} \citep{dfm2013},
\texttt{isoclassify} \citep{Huber2017},
\texttt{Jupyter} \citep{jupyter2016},
\texttt{KeckSpec} \citep{Rice2020}
\texttt{matplotlib} \citep{hunter2007},
\texttt{numpy} \citep{vanderwalt2011},
\texttt{pandas} \citep{pandas2010},
\texttt{Python Limb Darkening Toolkit} \citep{ldtk2015}
\texttt{RadVel} \citep{radvel},
\texttt{Smint} \citep{smint}
\texttt{SpecMatch-Syn} \citep{CKS1}
\texttt{Transit Least Squares} \citep{TLS}
\texttt{exoplanet} \citep{exoplanet} and its dependencies \citep{exoplanet:agol19, exoplanet:astropy18, exoplanet:espinoza18, exoplanet:luger18, exoplanet:pymc3, exoplanet:theano}

\section{Acknowledgments}

\par We thank the anonymous referee for their insightful and thorough comments. We are grateful to Tim Brandt for his insight and contributions to the methods of \S \ref{5thcomp}. We thank the time assignment committees of the University of California, the California Institute of Technology, NASA, and the University of Hawai`i for supporting the TESS-Keck Survey with observing time at Keck Observatory and on the Automated Planet Finder.  We thank NASA for funding associated with our Key Strategic Mission Support project.  We gratefully acknowledge the efforts and dedication of the Keck Observatory staff for support of HIRES and remote observing.  We recognize and acknowledge the cultural role and reverence that the summit of Maunakea has within the indigenous Hawaiian community. We are deeply grateful to have the opportunity to conduct observations from this mountain.  We thank Ken and Gloria Levy, who supported the construction of the Levy Spectrometer on the Automated Planet Finder. We thank the University of California and Google for supporting Lick Observatory and the UCO staff for their dedicated work scheduling and operating the telescopes of Lick Observatory. This paper is based on data collected by the TESS mission. Funding for the TESS mission is provided by the NASA Explorer Program. We acknowledge the use of public TESS data from pipelines at the TESS Science Office and at the TESS Science Processing Operations Center. This paper includes data collected by the TESS mission that are publicly available from the Mikulski Archive for Space Telescopes (MAST).

\par E.A.P. acknowledges the support of the Alfred P. Sloan Foundation. L.M.W. is supported by the Beatrice Watson Parrent Fellowship and NASA ADAP Grant 80NSSC19K0597. A.C. acknowledges support from the National Science Foundation through the Graduate Research Fellowship Program (DGE 1842402). D.H. acknowledges support from the Alfred P. Sloan Foundation, the National Aeronautics and Space Administration (80NSSC18K1585, 80NSSC19K0379), and the National Science Foundation (AST-1717000). I.J.M.C. acknowledges support from the NSF through grant AST-1824644. P.D. acknowledges support from a National Science Foundation Astronomy and Astrophysics Postdoctoral Fellowship under award AST-1903811. A.B. is supported by the NSF Graduate Research Fellowship, grant No. DGE 1745301. R.A.R. is supported by the NSF Graduate Research Fellowship, grant No. DGE 1745301. C. D. D. acknowledges the support of the Hellman Family Faculty Fund, the Alfred P. Sloan Foundation, the David \& Lucile Packard Foundation, and the National Aeronautics and Space Administration via the TESS Guest Investigator Program (80NSSC18K1583). J.M.A.M. is supported by the NSF Graduate Research Fellowship, grant No. DGE-1842400. J.M.A.M. also acknowledges the LSSTC Data Science Fellowship Program, which is funded by LSSTC, NSF Cybertraining Grant No. 1829740, the Brinson Foundation, and the Moore Foundation; his participation in the program has benefited this work. M.R.K is supported by the NSF Graduate Research Fellowship, grant No. DGE 1339067.

\bibliography{bib.bib}

\begin{thebibliography}{}
\expandafter\ifx\csname natexlab\endcsname\relax\def\natexlab#1{#1}\fi
\providecommand{\url}[1]{\href{#1}{#1}}
\providecommand{\dodoi}[1]{doi:~\href{http://doi.org/#1}{\nolinkurl{#1}}}
\providecommand{\doeprint}[1]{\href{http://ascl.net/#1}{\nolinkurl{http://ascl.net/#1}}}
\providecommand{\doarXiv}[1]{\href{https://arxiv.org/abs/#1}{\nolinkurl{https://arxiv.org/abs/#1}}}

\bibitem[{Hip(1997)}]{HipparcosCatalog}
 1997, ESA Special Publication, Vol. 1200, {The HIPPARCOS and TYCHO catalogues.
  Astrometric and photometric star catalogues derived from the ESA HIPPARCOS
  Space Astrometry Mission}

\bibitem[{{Agol} {et~al.}(2019){Agol}, {Luger}, \&
  {Foreman-Mackey}}]{exoplanet:agol19}
{Agol}, E., {Luger}, R., \& {Foreman-Mackey}, D. 2019, arXiv e-prints

\bibitem[{{Akaike}(1974)}]{Akaike1974}
{Akaike}, H. 1974, IEEE Transactions on Automatic Control, 19, 716,
  \dodoi{10.1109/TAC.1974.1100705}

\bibitem[{{Astropy Collaboration} {et~al.}(2013){Astropy Collaboration},
  {Robitaille}, {Tollerud}, {Greenfield}, {Droettboom}, {Bray}, {Aldcroft},
  {Davis}, {Ginsburg}, {Price-Whelan}, {Kerzendorf}, {Conley}, {Crighton},
  {Barbary}, {Muna}, {Ferguson}, {Grollier}, {Parikh}, {Nair}, {Unther},
  {Deil}, {Woillez}, {Conseil}, {Kramer}, {Turner}, {Singer}, {Fox}, {Weaver},
  {Zabalza}, {Edwards}, {Azalee Bostroem}, {Burke}, {Casey}, {Crawford},
  {Dencheva}, {Ely}, {Jenness}, {Labrie}, {Lim}, {Pierfederici}, {Pontzen},
  {Ptak}, {Refsdal}, {Servillat}, \& {Streicher}}]{astropy2013}
{Astropy Collaboration}, {Robitaille}, T.~P., {Tollerud}, E.~J., {et~al.} 2013,
  AAP, 558, A33, \dodoi{10.1051/0004-6361/201322068}

\bibitem[{{Astropy Collaboration} {et~al.}(2018){Astropy Collaboration},
  {Price-Whelan}, {Sip{\H o}cz}, {G{\"u}nther}, {Lim}, {Crawford}, {Conseil},
  {Shupe}, {Craig}, {Dencheva}, {Ginsburg}, {VanderPlas}, {Bradley},
  {P{\'e}rez-Su{\'a}rez}, {de Val-Borro}, {Aldcroft}, {Cruz}, {Robitaille},
  {Tollerud}, {Ardelean}, {Babej}, {Bach}, {Bachetti}, {Bakanov}, {Bamford},
  {Barentsen}, {Barmby}, {Baumbach}, {Berry}, {Biscani}, {Boquien}, {Bostroem},
  {Bouma}, {Brammer}, {Bray}, {Breytenbach}, {Buddelmeijer}, {Burke},
  {Calderone}, {Cano Rodr{\'{\i}}guez}, {Cara}, {Cardoso}, {Cheedella},
  {Copin}, {Corrales}, {Crichton}, {D'Avella}, {Deil}, {Depagne}, {Dietrich},
  {Donath}, {Droettboom}, {Earl}, {Erben}, {Fabbro}, {Ferreira}, {Finethy},
  {Fox}, {Garrison}, {Gibbons}, {Goldstein}, {Gommers}, {Greco}, {Greenfield},
  {Groener}, {Grollier}, {Hagen}, {Hirst}, {Homeier}, {Horton}, {Hosseinzadeh},
  {Hu}, {Hunkeler}, {Ivezi{\'c}}, {Jain}, {Jenness}, {Kanarek}, {Kendrew},
  {Kern}, {Kerzendorf}, {Khvalko}, {King}, {Kirkby}, {Kulkarni}, {Kumar},
  {Lee}, {Lenz}, {Littlefair}, {Ma}, {Macleod}, {Mastropietro}, {McCully},
  {Montagnac}, {Morris}, {Mueller}, {Mumford}, {Muna}, {Murphy}, {Nelson},
  {Nguyen}, {Ninan}, {N{\"o}the}, {Ogaz}, {Oh}, {Parejko}, {Parley}, {Pascual},
  {Patil}, {Patil}, {Plunkett}, {Prochaska}, {Rastogi}, {Reddy Janga},
  {Sabater}, {Sakurikar}, {Seifert}, {Sherbert}, {Sherwood-Taylor}, {Shih},
  {Sick}, {Silbiger}, {Singanamalla}, {Singer}, {Sladen}, {Sooley},
  {Sornarajah}, {Streicher}, {Teuben}, {Thomas}, {Tremblay}, {Turner},
  {Terr{\'o}n}, {van Kerkwijk}, {de la Vega}, {Watkins}, {Weaver}, {Whitmore},
  {Woillez}, {Zabalza}, \& {Astropy Contributors}}]{exoplanet:astropy18}
{Astropy Collaboration}, {Price-Whelan}, A.~M., {Sip{\H o}cz}, B.~M., {et~al.}
  2018, \aj, 156, 123, \dodoi{10.3847/1538-3881/aabc4f}

\bibitem[{{Badenas-Agusti} {et~al.}(2020){Badenas-Agusti}, {G{\"u}nther},
  {Daylan}, {Mikal-Evans}, {Vanderburg}, {Huang}, {Matthews}, {Rackham},
  {Bieryla}, {Stassun}, {Kane}, {Shporer}, {Fulton}, {Hill}, {Nowak}, {Ribas},
  {Pall{\'e}}, {Jenkins}, {Latham}, {Seager}, {Ricker}, {Vanderspek}, {Winn},
  {Abril-Pla}, {Collins}, {Guerra Serra}, {Niraula}, {Rustamkulov}, {Barclay},
  {Crossfield}, {Howell}, {Ciardi}, {Gonzales}, {Schlieder}, {Caldwell},
  {Fausnaugh}, {McDermott}, {Paegert}, {Pepper}, {Rose}, \& {Twicken}}]{BA2020}
{Badenas-Agusti}, M., {G{\"u}nther}, M.~N., {Daylan}, T., {et~al.} 2020, arXiv
  e-prints, arXiv:2002.03958.
\newblock \doarXiv{2002.03958}

\bibitem[{{Batalha} {et~al.}(2017){Batalha}, {Mandell}, {Pontoppidan},
  {Stevenson}, {Lewis}, {Kalirai}, {Earl}, {Greene}, {Albert}, \&
  {Nielsen}}]{batalha17}
{Batalha}, N.~E., {Mandell}, A., {Pontoppidan}, K., {et~al.} 2017, \pasp, 129,
  064501, \dodoi{10.1088/1538-3873/aa65b0}

\bibitem[{{Batygin} \& {Adams}(2017)}]{Batygin2017}
{Batygin}, K., \& {Adams}, F.~C. 2017, \aj, 153, 120,
  \dodoi{10.3847/1538-3881/153/3/120}

\bibitem[{{Berger} {et~al.}(2020){Berger}, {Huber}, {van Saders}, {Gaidos},
  {Tayar}, \& {Kraus}}]{Berger2020}
{Berger}, T.~A., {Huber}, D., {van Saders}, J.~L., {et~al.} 2020, \aj, 159,
  280, \dodoi{10.3847/1538-3881/159/6/280}

\bibitem[{{Boss}(1997)}]{Boss97}
{Boss}, A.~P. 1997, Science, 276, 1836, \dodoi{10.1126/science.276.5320.1836}

\bibitem[{{Bou{\'e}} {et~al.}(2013){Bou{\'e}}, {Montalto}, {Boisse}, {Oshagh},
  \& {Santos}}]{Boue2013}
{Bou{\'e}}, G., {Montalto}, M., {Boisse}, I., {Oshagh}, M., \& {Santos}, N.~C.
  2013, \aap, 550, A53, \dodoi{10.1051/0004-6361/201220146}

\bibitem[{{Brandt}(2021)}]{Brandt2021}
{Brandt}, T.~D. 2021, arXiv e-prints, arXiv:2105.11662.
\newblock \doarXiv{2105.11662}

\bibitem[{{Brewer} \& {Fischer}(2017)}]{Brewer2016}
{Brewer}, J.~M., \& {Fischer}, D.~A. 2017, \apj, 840, 121,
  \dodoi{10.3847/1538-4357/aa6d53}

\bibitem[{{Bryan} {et~al.}(2019){Bryan}, {Knutson}, {Lee}, {Fulton}, {Batygin},
  {Ngo}, \& {Meshkat}}]{Bryan2019}
{Bryan}, M.~L., {Knutson}, H.~A., {Lee}, E.~J., {et~al.} 2019, \aj, 157, 52,
  \dodoi{10.3847/1538-3881/aaf57f}

\bibitem[{{Butler} {et~al.}(1996){Butler}, {Marcy}, {Williams}, {McCarthy},
  {Dosanjh}, \& {Vogt}}]{Butler1996}
{Butler}, R.~P., {Marcy}, G.~W., {Williams}, E., {et~al.} 1996, \pasp, 108,
  500, \dodoi{10.1086/133755}

\bibitem[{{Carter} {et~al.}(2012){Carter}, {Agol}, {Chaplin}, {Basu},
  {Bedding}, {Buchhave}, {Christensen-Dalsgaard}, {Deck}, {Elsworth},
  {Fabrycky}, {Ford}, {Fortney}, {Hale}, {Handberg}, {Hekker}, {Holman},
  {Huber}, {Karoff}, {Kawaler}, {Kjeldsen}, {Lissauer}, {Lopez}, {Lund},
  {Lundkvist}, {Metcalfe}, {Miglio}, {Rogers}, {Stello}, {Borucki}, {Bryson},
  {Christiansen}, {Cochran}, {Geary}, {Gilliland}, {Haas}, {Hall}, {Howard},
  {Jenkins}, {Klaus}, {Koch}, {Latham}, {MacQueen}, {Sasselov}, {Steffen},
  {Twicken}, \& {Winn}}]{Carter2012}
{Carter}, J.~A., {Agol}, E., {Chaplin}, W.~J., {et~al.} 2012, Science, 337,
  556, \dodoi{10.1126/science.1223269}

\bibitem[{{Choi} {et~al.}(2016){Choi}, {Dotter}, {Conroy}, {Cantiello},
  {Paxton}, \& {Johnson}}]{Choi2016}
{Choi}, J., {Dotter}, A., {Conroy}, C., {et~al.} 2016, \apj, 823, 102,
  \dodoi{10.3847/0004-637X/823/2/102}

\bibitem[{{Chontos} {et~al.}(2021){Chontos}, {Akana Murphy}, {MacDougall},
  {Fetherolf}, {Van Zandt}, {Rubenzahl}, {Beard}, {Huber}, {Batalha},
  {Crossfield}, {Dressing}, {Fulton}, {Howard}, {Isaacson}, {Kane}, {Petigura},
  {Robertson}, {Roy}, {Weiss}, {Behmard}, {Dai}, {Dalba}, {Giacalone}, {Hill},
  {Lubin}, {Mayo}, {Mocnik}, {Polanski}, {Rosenthal}, {Scarsdale}, \&
  {Turtelboom}}]{Chontos2021}
{Chontos}, A., {Akana Murphy}, J.~M., {MacDougall}, M.~G., {et~al.} 2021, arXiv
  e-prints, arXiv:2106.06156.
\newblock \doarXiv{2106.06156}

\bibitem[{{Dai} {et~al.}(2020){Dai}, {Roy}, {Fulton}, {Robertson}, {Hirsch},
  {Isaacson}, {Albrecht}, {Mann}, {Kristiansen}, {Batalha}, {Beard}, {Behmard},
  {Chontos}, {Crossfield}, {Dalba}, {Dressing}, {Giacalone}, {Hill}, {Howard},
  {Huber}, {Kane}, {Kosiarek}, {Lubin}, {Mayo}, {Mocnik}, {Akana Murphy},
  {Petigura}, {Rosenthal}, {Rubenzahl}, {Scarsdale}, {Weiss}, {Van Zandt},
  {Ricker}, {Vanderspek}, {Latham}, {Seager}, {Winn}, {Jenkins}, {Caldwell},
  {Charbonneau}, {Daylan}, {G{\"u}nther}, {Morgan}, {Quinn}, {Rose}, \&
  {Smith}}]{Dai2020}
{Dai}, F., {Roy}, A., {Fulton}, B., {et~al.} 2020, \aj, 160, 193,
  \dodoi{10.3847/1538-3881/abb3bd}

\bibitem[{{Dalba} {et~al.}(2020){Dalba}, {Gupta}, {Rodriguez}, {Dragomir},
  {Huang}, {Kane}, {Quinn}, {Bieryla}, {Esquerdo}, {Fulton}, {Scarsdale},
  {Batalha}, {Beard}, {Behmard}, {Chontos}, {Crossfield}, {Dressing},
  {Giacalone}, {Hill}, {Hirsch}, {Howard}, {Huber}, {Isaacson}, {Kosiarek},
  {Lubin}, {Mayo}, {Mocnik}, {Akana Murphy}, {Petigura}, {Robertson},
  {Rosenthal}, {Roy}, {Rubenzahl}, {Van Zandt}, {Weiss}, {Knudstrup},
  {Andersen}, {Grundahl}, {Yao}, {Pepper}, {Villanueva}, {Ciardi}, {Cloutier},
  {Jacobs}, {Kristiansen}, {LaCourse}, {Lendl}, {Osborn}, {Palle}, {Stassun},
  {Stevens}, {Ricker}, {Vanderspek}, {Latham}, {Seager}, {Winn}, {Jenkins},
  {Caldwell}, {Daylan}, {Fong}, {Goeke}, {Rose}, {Rowden}, {Schlieder},
  {Smith}, \& {Vanderburg}}]{Dalba2020}
{Dalba}, P.~A., {Gupta}, A.~F., {Rodriguez}, J.~E., {et~al.} 2020, \aj, 159,
  241, \dodoi{10.3847/1538-3881/ab84e3}

\bibitem[{{Espinoza}(2018)}]{exoplanet:espinoza18}
{Espinoza}, N. 2018, Research Notes of the American Astronomical Society, 2,
  209, \dodoi{10.3847/2515-5172/aaef38}

\bibitem[{Foreman-Mackey(2016)}]{dfm2016}
Foreman-Mackey, D. 2016, The Journal of Open Source Software, 24,
  \dodoi{10.21105/joss.00024}

\bibitem[{{Foreman-Mackey} {et~al.}(2013){Foreman-Mackey}, {Hogg}, {Lang}, \&
  {Goodman}}]{dfm2013}
{Foreman-Mackey}, D., {Hogg}, D.~W., {Lang}, D., \& {Goodman}, J. 2013, PASP,
  125, 306, \dodoi{10.1086/670067}

\bibitem[{Foreman-Mackey {et~al.}(2020{\natexlab{a}})Foreman-Mackey, Luger,
  Czekala, Agol, Price-Whelan, \& Barclay}]{exoplanet}
Foreman-Mackey, D., Luger, R., Czekala, I., {et~al.} 2020{\natexlab{a}},
  exoplanet-dev/exoplanet v0.3.2, \dodoi{10.5281/zenodo.1998447}

\bibitem[{Foreman-Mackey {et~al.}(2020{\natexlab{b}})Foreman-Mackey, Luger,
  Czekala, Agol, Price-Whelan, Gilbert, Brandt, Barclay, \&
  Bouma}]{exoplanetDFM}
---. 2020{\natexlab{b}}, exoplanet-dev/exoplanet v0.4.1,
  \dodoi{10.5281/zenodo.1998447}

\bibitem[{{Fulton} \& {Petigura}(2018)}]{FulPet2018}
{Fulton}, B.~J., \& {Petigura}, E.~A. 2018, \aj, 156, 264,
  \dodoi{10.3847/1538-3881/aae828}

\bibitem[{{Fulton} {et~al.}(2018){Fulton}, {Petigura}, {Blunt}, \&
  {Sinukoff}}]{radvel}
{Fulton}, B.~J., {Petigura}, E.~A., {Blunt}, S., \& {Sinukoff}, E. 2018, \pasp,
  130, 044504, \dodoi{10.1088/1538-3873/aaaaa8}

\bibitem[{{Fulton} {et~al.}(2017){Fulton}, {Petigura}, {Howard}, {Isaacson},
  {Marcy}, {Cargile}, {Hebb}, {Weiss}, {Johnson}, {Morton}, {Sinukoff},
  {Crossfield}, \& {Hirsch}}]{Fulton2017}
{Fulton}, B.~J., {Petigura}, E.~A., {Howard}, A.~W., {et~al.} 2017, \aj, 154,
  109, \dodoi{10.3847/1538-3881/aa80eb}

\bibitem[{{Fulton} {et~al.}(2021){Fulton}, {Rosenthal}, {Hirsch}, {Isaacson},
  {Howard}, {Dedrick}, {Sherstyuk}, {Blunt}, {Petigura}, {Knutson}, {Behmard},
  {Chontos}, {Crepp}, {Crossfield}, {Dalba}, {Fischer}, {Henry}, {Kane},
  {Kosiarek}, {Marcy}, {Rubenzahl}, {Weiss}, \& {Wright}}]{Fulton2021}
{Fulton}, B.~J., {Rosenthal}, L.~J., {Hirsch}, L.~A., {et~al.} 2021, arXiv
  e-prints, arXiv:2105.11584.
\newblock \doarXiv{2105.11584}

\bibitem[{{Gaia Collaboration} {et~al.}(2018){Gaia Collaboration}, {Brown},
  {Vallenari}, {Prusti}, {de Bruijne}, {Babusiaux}, {Bailer-Jones}, {Biermann},
  {Evans}, {Eyer}, {Jansen}, {Jordi}, {Klioner}, {Lammers}, {Lindegren},
  {Luri}, {Mignard}, {Panem}, {Pourbaix}, {Randich}, {Sartoretti}, {Siddiqui},
  {Soubiran}, {van Leeuwen}, {Walton}, {Arenou}, {Bastian}, {Cropper},
  {Drimmel}, {Katz}, {Lattanzi}, {Bakker}, {Cacciari}, {Casta{\~n}eda},
  {Chaoul}, {Cheek}, {De Angeli}, {Fabricius}, {Guerra}, {Holl}, {Masana},
  {Messineo}, {Mowlavi}, {Nienartowicz}, {Panuzzo}, {Portell}, {Riello},
  {Seabroke}, {Tanga}, {Th{\'e}venin}, {Gracia-Abril}, {Comoretto},
  {Garcia-Reinaldos}, {Teyssier}, {Altmann}, {Andrae}, {Audard},
  {Bellas-Velidis}, {Benson}, {Berthier}, {Blomme}, {Burgess}, {Busso},
  {Carry}, {Cellino}, {Clementini}, {Clotet}, {Creevey}, {Davidson}, {De
  Ridder}, {Delchambre}, {Dell'Oro}, {Ducourant},
  {Fern{\'a}ndez-Hern{\'a}ndez}, {Fouesneau}, {Fr{\'e}mat}, {Galluccio},
  {Garc{\'\i}a-Torres}, {Gonz{\'a}lez-N{\'u}{\~n}ez}, {Gonz{\'a}lez-Vidal},
  {Gosset}, {Guy}, {Halbwachs}, {Hambly}, {Harrison}, {Hern{\'a}ndez},
  {Hestroffer}, {Hodgkin}, {Hutton}, {Jasniewicz}, {Jean-Antoine-Piccolo},
  {Jordan}, {Korn}, {Krone-Martins}, {Lanzafame}, {Lebzelter}, {L{\"o}ffler},
  {Manteiga}, {Marrese}, {Mart{\'\i}n-Fleitas}, {Moitinho}, {Mora}, {Muinonen},
  {Osinde}, {Pancino}, {Pauwels}, {Petit}, {Recio-Blanco}, {Richards},
  {Rimoldini}, {Robin}, {Sarro}, {Siopis}, {Smith}, {Sozzetti}, {S{\"u}veges},
  {Torra}, {van Reeven}, {Abbas}, {Abreu Aramburu}, {Accart}, {Aerts},
  {Altavilla}, {{\'A}lvarez}, {Alvarez}, {Alves}, {Anderson}, {Andrei},
  {Anglada Varela}, {Antiche}, {Antoja}, {Arcay}, {Astraatmadja}, {Bach},
  {Baker}, {Balaguer-N{\'u}{\~n}ez}, {Balm}, {Barache}, {Barata}, {Barbato},
  {Barblan}, {Barklem}, {Barrado}, {Barros}, {Barstow}, {Bartholom{\'e}
  Mu{\~n}oz}, {Bassilana}, {Becciani}, {Bellazzini}, {Berihuete}, {Bertone},
  {Bianchi}, {Bienaym{\'e}}, {Blanco-Cuaresma}, {Boch}, {Boeche}, {Bombrun},
  {Borrachero}, {Bossini}, {Bouquillon}, {Bourda}, {Bragaglia}, {Bramante},
  {Breddels}, {Bressan}, {Brouillet}, {Br{\"u}semeister}, {Brugaletta},
  {Bucciarelli}, {Burlacu}, {Busonero}, {Butkevich}, {Buzzi}, {Caffau},
  {Cancelliere}, {Cannizzaro}, {Cantat-Gaudin}, {Carballo}, {Carlucci},
  {Carrasco}, {Casamiquela}, {Castellani}, {Castro-Ginard}, {Charlot},
  {Chemin}, {Chiavassa}, {Cocozza}, {Costigan}, {Cowell}, {Crifo}, {Crosta},
  {Crowley}, {Cuypers}, {Dafonte}, {Damerdji}, {Dapergolas}, {David}, {David},
  {de Laverny}, {De Luise}, {De March}, {de Martino}, {de Souza}, {de Torres},
  {Debosscher}, {del Pozo}, {Delbo}, {Delgado}, {Delgado}, {Di Matteo},
  {Diakite}, {Diener}, {Distefano}, {Dolding}, {Drazinos}, {Dur{\'a}n},
  {Edvardsson}, {Enke}, {Eriksson}, {Esquej}, {Eynard Bontemps}, {Fabre},
  {Fabrizio}, {Faigler}, {Falc{\~a}o}, {Farr{\`a}s Casas}, {Federici},
  {Fedorets}, {Fernique}, {Figueras}, {Filippi}, {Findeisen}, {Fonti},
  {Fraile}, {Fraser}, {Fr{\'e}zouls}, {Gai}, {Galleti}, {Garabato},
  {Garc{\'\i}a-Sedano}, {Garofalo}, {Garralda}, {Gavel}, {Gavras}, {Gerssen},
  {Geyer}, {Giacobbe}, {Gilmore}, {Girona}, {Giuffrida}, {Glass}, {Gomes},
  {Granvik}, {Gueguen}, {Guerrier}, {Guiraud}, {Guti{\'e}rrez-S{\'a}nchez},
  {Haigron}, {Hatzidimitriou}, {Hauser}, {Haywood}, {Heiter}, {Helmi}, {Heu},
  {Hilger}, {Hobbs}, {Hofmann}, {Holland}, {Huckle}, {Hypki}, {Icardi},
  {Jan{\ss}en}, {Jevardat de Fombelle}, {Jonker}, {Juh{\'a}sz}, {Julbe},
  {Karampelas}, {Kewley}, {Klar}, {Kochoska}, {Kohley}, {Kolenberg},
  {Kontizas}, {Kontizas}, {Koposov}, {Kordopatis}, {Kostrzewa-Rutkowska},
  {Koubsky}, {Lambert}, {Lanza}, {Lasne}, {Lavigne}, {Le Fustec}, {Le
  Poncin-Lafitte}, {Lebreton}, {Leccia}, {Leclerc}, {Lecoeur-Taibi},
  {Lenhardt}, {Leroux}, {Liao}, {Licata}, {Lindstr{\o}m}, {Lister}, {Livanou},
  {Lobel}, {L{\'o}pez}, {Managau}, {Mann}, {Mantelet}, {Marchal}, {Marchant},
  {Marconi}, {Marinoni}, {Marschalk{\'o}}, {Marshall}, {Martino}, {Marton},
  {Mary}, {Massari}, {Matijevi{\v{c}}}, {Mazeh}, {McMillan}, {Messina},
  {Michalik}, {Millar}, {Molina}, {Molinaro}, {Moln{\'a}r}, {Montegriffo},
  {Mor}, {Morbidelli}, {Morel}, {Morris}, {Mulone}, {Muraveva}, {Musella},
  {Nelemans}, {Nicastro}, {Noval}, {O'Mullane}, {Ord{\'e}novic},
  {Ord{\'o}{\~n}ez-Blanco}, {Osborne}, {Pagani}, {Pagano}, {Pailler},
  {Palacin}, {Palaversa}, {Panahi}, {Pawlak}, {Piersimoni}, {Pineau}, {Plachy},
  {Plum}, {Poggio}, {Poujoulet}, {Pr{\v{s}}a}, {Pulone}, {Racero}, {Ragaini},
  {Rambaux}, {Ramos-Lerate}, {Regibo}, {Reyl{\'e}}, {Riclet}, {Ripepi}, {Riva},
  {Rivard}, {Rixon}, {Roegiers}, {Roelens}, {Romero-G{\'o}mez}, {Rowell},
  {Royer}, {Ruiz-Dern}, {Sadowski}, {Sagrist{\`a} Sell{\'e}s}, {Sahlmann},
  {Salgado}, {Salguero}, {Sanna}, {Santana-Ros}, {Sarasso}, {Savietto},
  {Schultheis}, {Sciacca}, {Segol}, {Segovia}, {S{\'e}gransan}, {Shih},
  {Siltala}, {Silva}, {Smart}, {Smith}, {Solano}, {Solitro}, {Sordo}, {Soria
  Nieto}, {Souchay}, {Spagna}, {Spoto}, {Stampa}, {Steele},
  {Steidelm{\"u}ller}, {Stephenson}, {Stoev}, {Suess}, {Surdej}, {Szabados},
  {Szegedi-Elek}, {Tapiador}, {Taris}, {Tauran}, {Taylor}, {Teixeira},
  {Terrett}, {Teyssand ier}, {Thuillot}, {Titarenko}, {Torra Clotet}, {Turon},
  {Ulla}, {Utrilla}, {Uzzi}, {Vaillant}, {Valentini}, {Valette}, {van Elteren},
  {Van Hemelryck}, {van Leeuwen}, {Vaschetto}, {Vecchiato}, {Veljanoski},
  {Viala}, {Vicente}, {Vogt}, {von Essen}, {Voss}, {Votruba}, {Voutsinas},
  {Walmsley}, {Weiler}, {Wertz}, {Wevers}, {Wyrzykowski}, {Yoldas},
  {{\v{Z}}erjal}, {Ziaeepour}, {Zorec}, {Zschocke}, {Zucker}, {Zurbach}, \&
  {Zwitter}}]{GaiaCollab}
{Gaia Collaboration}, {Brown}, A.~G.~A., {Vallenari}, A., {et~al.} 2018, \aap,
  616, A1, \dodoi{10.1051/0004-6361/201833051}

\bibitem[{{Gaudi} \& {Winn}(2007)}]{GaudiWinn2007}
{Gaudi}, B.~S., \& {Winn}, J.~N. 2007, \apj, 655, 550, \dodoi{10.1086/509910}

\bibitem[{{Gilbert} \& {Fabrycky}(2020)}]{GilbertandFabrycky2020}
{Gilbert}, G.~J., \& {Fabrycky}, D.~C. 2020, \aj, 159, 281,
  \dodoi{10.3847/1538-3881/ab8e3c}

\bibitem[{{Hippke} \& {Heller}(2019)}]{TLS}
{Hippke}, M., \& {Heller}, R. 2019, \aap, 623, A39,
  \dodoi{10.1051/0004-6361/201834672}

\bibitem[{{Howard} {et~al.}(2010){Howard}, {Johnson}, {Marcy}, {Fischer},
  {Wright}, {Bernat}, {Henry}, {Peek}, {Isaacson}, {Apps}, {Endl}, {Cochran},
  {Valenti}, {Anderson}, \& {Piskunov}}]{Howard2010}
{Howard}, A.~W., {Johnson}, J.~A., {Marcy}, G.~W., {et~al.} 2010, \apj, 721,
  1467, \dodoi{10.1088/0004-637X/721/2/1467}

\bibitem[{{Howard} {et~al.}(2012){Howard}, {Marcy}, {Bryson}, {Jenkins},
  {Rowe}, {Batalha}, {Borucki}, {Koch}, {Dunham}, {Gautier}, {Van Cleve},
  {Cochran}, {Latham}, {Lissauer}, {Torres}, {Brown}, {Gilliland}, {Buchhave},
  {Caldwell}, {Christensen-Dalsgaard}, {Ciardi}, {Fressin}, {Haas}, {Howell},
  {Kjeldsen}, {Seager}, {Rogers}, {Sasselov}, {Steffen}, {Basri},
  {Charbonneau}, {Christiansen}, {Clarke}, {Dupree}, {Fabrycky}, {Fischer},
  {Ford}, {Fortney}, {Tarter}, {Girouard}, {Holman}, {Johnson}, {Klaus},
  {Machalek}, {Moorhead}, {Morehead}, {Ragozzine}, {Tenenbaum}, {Twicken},
  {Quinn}, {Isaacson}, {Shporer}, {Lucas}, {Walkowicz}, {Welsh}, {Boss},
  {Devore}, {Gould}, {Smith}, {Morris}, {Prsa}, {Morton}, {Still}, {Thompson},
  {Mullally}, {Endl}, \& {MacQueen}}]{Howard2012}
{Howard}, A.~W., {Marcy}, G.~W., {Bryson}, S.~T., {et~al.} 2012, \apjs, 201,
  15, \dodoi{10.1088/0067-0049/201/2/15}

\bibitem[{{Huber} {et~al.}(2017){Huber}, {Zinn}, {Bojsen-Hansen},
  {Pinsonneault}, {Sahlholdt}, {Serenelli}, {Silva Aguirre}, {Stassun},
  {Stello}, {Tayar}, {Bastien}, {Bedding}, {Buchhave}, {Chaplin}, {Davies},
  {Garc{\'\i}a}, {Latham}, {Mathur}, {Mosser}, \& {Sharma}}]{Huber2017}
{Huber}, D., {Zinn}, J., {Bojsen-Hansen}, M., {et~al.} 2017, \apj, 844, 102,
  \dodoi{10.3847/1538-4357/aa75ca}

\bibitem[{{Hunter}(2007)}]{hunter2007}
{Hunter}, J.~D. 2007, Computing in Science and Engineering, 9, 90,
  \dodoi{10.1109/MCSE.2007.55}

\bibitem[{{Isaacson} \& {Fischer}(2010)}]{IandF2010}
{Isaacson}, H., \& {Fischer}, D. 2010, \apj, 725, 875,
  \dodoi{10.1088/0004-637X/725/1/875}

\bibitem[{{Jenkins} {et~al.}(2016){Jenkins}, {Twicken}, {McCauliff},
  {Campbell}, {Sanderfer}, {Lung}, {Mansouri-Samani}, {Girouard}, {Tenenbaum},
  {Klaus}, {Smith}, {Caldwell}, {Chacon}, {Henze}, {Heiges}, {Latham},
  {Morgan}, {Swade}, {Rinehart}, \& {Vanderspek}}]{Jenkins2016}
{Jenkins}, J.~M., {Twicken}, J.~D., {McCauliff}, S., {et~al.} 2016, in Society
  of Photo-Optical Instrumentation Engineers (SPIE) Conference Series, Vol.
  9913, Software and Cyberinfrastructure for Astronomy IV, ed. G.~{Chiozzi} \&
  J.~C. {Guzman}, 99133E, \dodoi{10.1117/12.2233418}

\bibitem[{{Jontof-Hutter} {et~al.}(2014){Jontof-Hutter}, {Lissauer}, {Rowe}, \&
  {Fabrycky}}]{JontofHutter2014}
{Jontof-Hutter}, D., {Lissauer}, J.~J., {Rowe}, J.~F., \& {Fabrycky}, D.~C.
  2014, \apj, 785, 15, \dodoi{10.1088/0004-637X/785/1/15}

\bibitem[{{Kempton} {et~al.}(2018){Kempton}, {Bean}, {Louie}, {Deming}, {Koll},
  {Mansfield}, {Christiansen}, {L{\'o}pez-Morales}, {Swain}, {Zellem},
  {Ballard}, {Barclay}, {Barstow}, {Batalha}, {Beatty}, {Berta-Thompson},
  {Birkby}, {Buchhave}, {Charbonneau}, {Cowan}, {Crossfield}, {de Val-Borro},
  {Doyon}, {Dragomir}, {Gaidos}, {Heng}, {Hu}, {Kane}, {Kreidberg}, {Mallonn},
  {Morley}, {Narita}, {Nascimbeni}, {Pall{\'e}}, {Quintana}, {Rauscher},
  {Seager}, {Shkolnik}, {Sing}, {Sozzetti}, {Stassun}, {Valenti}, \& {von
  Essen}}]{Kempton2018}
{Kempton}, E. M.~R., {Bean}, J.~L., {Louie}, D.~R., {et~al.} 2018, \pasp, 130,
  114401, \dodoi{10.1088/1538-3873/aadf6f}

\bibitem[{{Kipping}(2013)}]{kipping2013}
{Kipping}, D.~M. 2013, \mnras, 434, L51, \dodoi{10.1093/mnrasl/slt075}

\bibitem[{Kluyver {et~al.}(2016)Kluyver, Ragan-Kelley, P{\'e}rez, Granger,
  Bussonnier, Frederic, Kelley, Hamrick, Grout, Corlay, Ivanov, Avila, Abdalla,
  Willing, \& development team}]{jupyter2016}
Kluyver, T., Ragan-Kelley, B., P{\'e}rez, F., {et~al.} 2016, in Positioning and
  Power in Academic Publishing: Players, Agents and Agendas, ed. F.~Loizides \&
  B.~Scmidt (IOS Press), 87--90.
\newblock \url{https://eprints.soton.ac.uk/403913/}

\bibitem[{{Lee} \& {Chiang}(2016)}]{LeeChiang2016}
{Lee}, E.~J., \& {Chiang}, E. 2016, \apj, 817, 90,
  \dodoi{10.3847/0004-637X/817/2/90}

\bibitem[{{Leleu} {et~al.}(2021){Leleu}, {Alibert}, {Hara}, {Hooton}, {Wilson},
  {Robutel}, {Delisle}, {Laskar}, {Hoyer}, {Lovis}, {Bryant}, {Ducrot},
  {Cabrera}, {Delrez}, {Acton}, {Adibekyan}, {Allart}, {Allende Prieto},
  {Alonso}, {Alves}, {Anderson}, {Angerhausen}, {Anglada Escud{\'e}},
  {Asquier}, {Barrado}, {Barros}, {Baumjohann}, {Bayliss}, {Beck}, {Beck},
  {Bekkelien}, {Benz}, {Billot}, {Bonfanti}, {Bonfils}, {Bouchy}, {Bourrier},
  {Bou{\'e}}, {Brandeker}, {Broeg}, {Buder}, {Burdanov}, {Burleigh},
  {B{\'a}rczy}, {Cameron}, {Chamberlain}, {Charnoz}, {Cooke}, {Corral Van
  Damme}, {Correia}, {Cristiani}, {Damasso}, {Davies}, {Deleuil}, {Demangeon},
  {Demory}, {Di Marcantonio}, {Di Persio}, {Dumusque}, {Ehrenreich}, {Erikson},
  {Figueira}, {Fortier}, {Fossati}, {Fridlund}, {Futyan}, {Gandolfi},
  {Garc{\'\i}a Mu{\~n}oz}, {Garcia}, {Gill}, {Gillen}, {Gillon}, {Goad},
  {Gonz{\'a}lez Hern{\'a}ndez}, {Guedel}, {G{\"u}nther}, {Haldemann},
  {Henderson}, {Heng}, {Hogan}, {Isaak}, {Jehin}, {Jenkins}, {Jord{\'a}n},
  {Kiss}, {Kristiansen}, {Lam}, {Lavie}, {Lecavelier des Etangs}, {Lendl},
  {Lillo-Box}, {Lo Curto}, {Magrin}, {Martins}, {Maxted}, {McCormac}, {Mehner},
  {Micela}, {Molaro}, {Moyano}, {Murray}, {Nascimbeni}, {Nunes}, {Olofsson},
  {Osborn}, {Oshagh}, {Ottensamer}, {Pagano}, {Pall{\'e}}, {Pedersen}, {Pepe},
  {Persson}, {Peter}, {Piotto}, {Polenta}, {Pollacco}, {Poretti}, {Pozuelos},
  {Queloz}, {Ragazzoni}, {Rando}, {Ratti}, {Rauer}, {Raynard}, {Rebolo},
  {Reimers}, {Ribas}, {Santos}, {Scandariato}, {Schneider}, {Sebastian},
  {Sestovic}, {Simon}, {Smith}, {Sousa}, {Sozzetti}, {Steller}, {Su{\'a}rez
  Mascare{\~n}o}, {Szab{\'o}}, {S{\'e}gransan}, {Thomas}, {Thompson},
  {Tilbrook}, {Triaud}, {Turner}, {Udry}, {Van Grootel}, {Venus}, {Verrecchia},
  {Vines}, {Walton}, {West}, {Wheatley}, {Wolter}, \& {Zapatero
  Osorio}}]{Leleu2021}
{Leleu}, A., {Alibert}, Y., {Hara}, N.~C., {et~al.} 2021, arXiv e-prints,
  arXiv:2101.09260.
\newblock \doarXiv{2101.09260}

\bibitem[{{Lindegren} {et~al.}(2020){Lindegren}, {Klioner}, {Hern{\'a}ndez},
  {Bombrun}, {Ramos-Lerate}, {Steidelm{\"u}ller}, {Bastian}, {Biermann}, {de
  Torres}, {Gerlach}, {Geyer}, {Hilger}, {Hobbs}, {Lammers}, {McMillan},
  {Stephenson}, {Casta{\~n}eda}, {Davidson}, {Fabricius}, {Gracia-Abril},
  {Portell}, {Rowell}, {Teyssier}, {Torra}, {Bartolom{\'e}}, {Clotet},
  {Garralda}, {Gonz{\'a}lez-Vidal}, {Torra}, {Abbas}, {Altmann}, {Anglada
  Varela}, {Balaguer-N{\'u}{\~n}ez}, {Balog}, {Barache}, {Becciani}, {Bernet},
  {Bertone}, {Bianchi}, {Bouquillon}, {Brown}, {Bucciarelli}, {Busonero},
  {Butkevich}, {Buzzi}, {Cancelliere}, {Carlucci}, {Charlot}, {Cioni},
  {Crosta}, {Crowley}, {del Peloso}, {del Pozo}, {Drimmel}, {Esquej}, {Fienga},
  {Fraile}, {Gai}, {Garcia-Reinaldos}, {Guerra}, {Hambly}, {Hauser},
  {Jan{\ss}en}, {Jordan}, {Kostrzewa-Rutkowska}, {Lattanzi}, {Liao}, {Licata},
  {Lister}, {L{\"o}ffler}, {Marchant}, {Masip}, {Mignard}, {Mints}, {Molina},
  {Mora}, {Morbidelli}, {Murphy}, {Pagani}, {Panuzzo}, {Pe{\~n}alosa Esteller},
  {Poggio}, {Re Fiorentin}, {Riva}, {Sagrist{\`a} Sell{\'e}s}, {Sanchez
  Gimenez}, {Sarasso}, {Sciacca}, {Siddiqui}, {Smart}, {Souami}, {Spagna},
  {Steele}, {Taris}, {Utrilla}, {van Reeven}, \& {Vecchiato}}]{GaiaEDR3}
{Lindegren}, L., {Klioner}, S.~A., {Hern{\'a}ndez}, J., {et~al.} 2020, arXiv
  e-prints, arXiv:2012.03380.
\newblock \doarXiv{2012.03380}

\bibitem[{{Lithwick} {et~al.}(2012){Lithwick}, {Xie}, \& {Wu}}]{Lithwick2012}
{Lithwick}, Y., {Xie}, J., \& {Wu}, Y. 2012, \apj, 761, 122,
  \dodoi{10.1088/0004-637X/761/2/122}

\bibitem[{{Lopez} \& {Fortney}(2014)}]{Lopez2014}
{Lopez}, E.~D., \& {Fortney}, J.~J. 2014, \apj, 792, 1,
  \dodoi{10.1088/0004-637X/792/1/1}

\bibitem[{{Luger} {et~al.}(2019){Luger}, {Agol}, {Foreman-Mackey}, {Fleming},
  {Lustig-Yaeger}, \& {Deitrick}}]{exoplanet:luger18}
{Luger}, R., {Agol}, E., {Foreman-Mackey}, D., {et~al.} 2019, \aj, 157, 64,
  \dodoi{10.3847/1538-3881/aae8e5}

\bibitem[{Marquis~de Laplace(1825)}]{LaplaceOriginal}
Marquis~de Laplace, P.-S. 1825, "Traite de Mecanique Celeste", Ch. VII,
  569–634

\bibitem[{{Masuda}(2014)}]{Masuda2014}
{Masuda}, K. 2014, \apj, 783, 53, \dodoi{10.1088/0004-637X/783/1/53}

\bibitem[{McKinney(2010)}]{pandas2010}
McKinney, W. 2010, in Proceedings of the 9th Python in Science Conference, ed.
  S.~van~der Walt \& J.~Millman, 51 -- 56

\bibitem[{{McLaughlin}(1924)}]{McLaughlin1924}
{McLaughlin}, D.~B. 1924, \apj, 60, 22, \dodoi{10.1086/142826}

\bibitem[{{Mikal-Evans} {et~al.}(2021){Mikal-Evans}, {Crossfield}, {Benneke},
  {Kreidberg}, {Moses}, {Morley}, {Thorngren}, {Molli{\`e}re},
  {Hardegree-Ullman}, {Brewer}, {Christiansen}, {Ciardi}, {Dragomir},
  {Dressing}, {Fortney}, {Gorjian}, {Greene}, {Hirsch}, {Howard}, {Howell},
  {Isaacson}, {Kosiarek}, {Krick}, {Livingston}, {Lothringer}, {Morales},
  {Petigura}, {Schlieder}, \& {Werner}}]{mikal-evans21}
{Mikal-Evans}, T., {Crossfield}, I. J.~M., {Benneke}, B., {et~al.} 2021, \aj,
  161, 18, \dodoi{10.3847/1538-3881/abc874}

\bibitem[{{Millholland}(2019)}]{Millholland2019}
{Millholland}, S. 2019, \apj, 886, 72, \dodoi{10.3847/1538-4357/ab4c3f}

\bibitem[{Millholland {et~al.}(2017)Millholland, Wang, \&
  Laughlin}]{Millholland2017}
Millholland, S., Wang, S., \& Laughlin, G. 2017, The Astrophysical Journal,
  849, L33, \dodoi{10.3847/2041-8213/aa9714}

\bibitem[{{Mills} {et~al.}(2019){Mills}, {Howard}, {Weiss}, {Steffen},
  {Isaacson}, {Fulton}, {Petigura}, {Kosiarek}, {Hirsch}, \&
  {Boisvert}}]{Mills2019}
{Mills}, S.~M., {Howard}, A.~W., {Weiss}, L.~M., {et~al.} 2019, \aj, 157, 145,
  \dodoi{10.3847/1538-3881/ab0899}

\bibitem[{Murray \& Dermott(2010)}]{murray_dermott_2010}
Murray, C.~D., \& Dermott, S.~F. 2010, Solar system dynamics (Cambridge Univ.
  Press)

\bibitem[{{Nissen} {et~al.}(2020){Nissen}, {Christensen-Dalsgaard},
  {Mosumgaard}, {Silva Aguirre}, {Spitoni}, \& {Verma}}]{Nissen2020}
{Nissen}, P.~E., {Christensen-Dalsgaard}, J., {Mosumgaard}, J.~R., {et~al.}
  2020, \aap, 640, A81, \dodoi{10.1051/0004-6361/202038300}

\bibitem[{{Ofir} {et~al.}(2014){Ofir}, {Dreizler}, {Zechmeister}, \&
  {Husser}}]{Ofir2014}
{Ofir}, A., {Dreizler}, S., {Zechmeister}, M., \& {Husser}, T.-O. 2014, \aap,
  561, A103, \dodoi{10.1051/0004-6361/201220935}

\bibitem[{{Owen} \& {Wu}(2017)}]{OwenWu2017}
{Owen}, J.~E., \& {Wu}, Y. 2017, \apj, 847, 29,
  \dodoi{10.3847/1538-4357/aa890a}

\bibitem[{Parviainen \& Aigrain(2015)}]{ldtk2015}
Parviainen, H., \& Aigrain, S. 2015, Monthly Notices of the Royal Astronomical
  Society, 453, 3821, \dodoi{10.1093/mnras/stv1857}

\bibitem[{{Pecaut} \& {Mamajek}(2013)}]{PecaultMamajek}
{Pecaut}, M.~J., \& {Mamajek}, E.~E. 2013, \apjs, 208, 9,
  \dodoi{10.1088/0067-0049/208/1/9}

\bibitem[{{Petigura} {et~al.}(2013){Petigura}, {Marcy}, \&
  {Howard}}]{Petigura2013}
{Petigura}, E.~A., {Marcy}, G.~W., \& {Howard}, A.~W. 2013, \apj, 770, 69,
  \dodoi{10.1088/0004-637X/770/1/69}

\bibitem[{{Petigura} {et~al.}(2017){Petigura}, {Howard}, {Marcy}, {Johnson},
  {Isaacson}, {Cargile}, {Hebb}, {Fulton}, {Weiss}, {Morton}, {Winn}, {Rogers},
  {Sinukoff}, {Hirsch}, \& {Crossfield}}]{CKS1}
{Petigura}, E.~A., {Howard}, A.~W., {Marcy}, G.~W., {et~al.} 2017, \aj, 154,
  107, \dodoi{10.3847/1538-3881/aa80de}

\bibitem[{Piaulet(2020)}]{smint}
Piaulet, C. 2020, Smint

\bibitem[{{Pollack} {et~al.}(1996){Pollack}, {Hubickyj}, {Bodenheimer},
  {Lissauer}, {Podolak}, \& {Greenzweig}}]{Pollack96}
{Pollack}, J.~B., {Hubickyj}, O., {Bodenheimer}, P., {et~al.} 1996, \icarus,
  124, 62, \dodoi{10.1006/icar.1996.0190}

\bibitem[{{Rice} \& {Brewer}(2020)}]{Rice2020}
{Rice}, M., \& {Brewer}, J.~M. 2020, \apj, 898, 119,
  \dodoi{10.3847/1538-4357/ab9f96}

\bibitem[{{Ricker} {et~al.}(2015){Ricker}, {Winn}, {Vanderspek}, {Latham},
  {Bakos}, {Bean}, {Berta-Thompson}, {Brown}, {Buchhave}, {Butler}, {Butler},
  {Chaplin}, {Charbonneau}, {Christensen-Dalsgaard}, {Clampin}, {Deming},
  {Doty}, {De Lee}, {Dressing}, {Dunham}, {Endl}, {Fressin}, {Ge}, {Henning},
  {Holman}, {Howard}, {Ida}, {Jenkins}, {Jernigan}, {Johnson}, {Kaltenegger},
  {Kawai}, {Kjeldsen}, {Laughlin}, {Levine}, {Lin}, {Lissauer}, {MacQueen},
  {Marcy}, {McCullough}, {Morton}, {Narita}, {Paegert}, {Palle}, {Pepe},
  {Pepper}, {Quirrenbach}, {Rinehart}, {Sasselov}, {Sato}, {Seager},
  {Sozzetti}, {Stassun}, {Sullivan}, {Szentgyorgyi}, {Torres}, {Udry}, \&
  {Villasenor}}]{ricker2015}
{Ricker}, G.~R., {Winn}, J.~N., {Vanderspek}, R., {et~al.} 2015, Journal of
  Astronomical Telescopes, Instruments, and Systems, 1, 014003,
  \dodoi{10.1117/1.JATIS.1.1.014003}

\bibitem[{{Rogers}(2015)}]{Rogers2015}
{Rogers}, L.~A. 2015, \apj, 801, 41, \dodoi{10.1088/0004-637X/801/1/41}

\bibitem[{{Rossiter}(1924)}]{Rossiter1924}
{Rossiter}, R.~A. 1924, \apj, 60, 15, \dodoi{10.1086/142825}

\bibitem[{{Rubenzahl} {et~al.}(2021){Rubenzahl}, {Dai}, {Howard}, {Chontos},
  {Giacalone}, {Lubin}, {Rosenthal}, {Isaacson}, {Batalha}, {Crossfield},
  {Dressing}, {Fulton}, {Huber}, {Kane}, {Petigura}, {Robertson}, {Roy},
  {Weiss}, {Beard}, {Hill}, {Mayo}, {Mo{\v{c}}nik}, {Akana Murphy}, \&
  {Scarsdale}}]{Rubenzhal2021}
{Rubenzahl}, R.~A., {Dai}, F., {Howard}, A.~W., {et~al.} 2021, arXiv e-prints,
  arXiv:2101.09371.
\newblock \doarXiv{2101.09371}

\bibitem[{Salvatier {et~al.}(2016)Salvatier, Wiecki, \&
  Fonnesbeck}]{exoplanet:pymc3}
Salvatier, J., Wiecki, T.~V., \& Fonnesbeck, C. 2016, PeerJ Computer Science,
  2, e55

\bibitem[{{Schlaufman}(2018)}]{Schlaufman2018}
{Schlaufman}, K.~C. 2018, \apj, 853, 37, \dodoi{10.3847/1538-4357/aa961c}

\bibitem[{{Stassun} {et~al.}(2018){Stassun}, {Oelkers}, {Pepper}, {Paegert},
  {De Lee}, {Torres}, {Latham}, {Charpinet}, {Dressing}, {Huber}, {Kane},
  {L{\'e}pine}, {Mann}, {Muirhead}, {Rojas-Ayala}, {Silvotti}, {Fleming},
  {Levine}, \& {Plavchan}}]{Stassun2018}
{Stassun}, K.~G., {Oelkers}, R.~J., {Pepper}, J., {et~al.} 2018, \aj, 156, 102,
  \dodoi{10.3847/1538-3881/aad050}

\bibitem[{{Tayar} {et~al.}(2020){Tayar}, {Claytor}, {Huber}, \& {van
  Saders}}]{Tayar2020}
{Tayar}, J., {Claytor}, Z.~R., {Huber}, D., \& {van Saders}, J. 2020, arXiv
  e-prints, arXiv:2012.07957.
\newblock \doarXiv{2012.07957}

\bibitem[{{Theano Development Team}(2016)}]{exoplanet:theano}
{Theano Development Team}. 2016, arXiv e-prints, abs/1605.02688.
\newblock \url{http://arxiv.org/abs/1605.02688}

\bibitem[{{Valenti} {et~al.}(1995){Valenti}, {Butler}, \&
  {Marcy}}]{Valenti1995}
{Valenti}, J.~A., {Butler}, R.~P., \& {Marcy}, G.~W. 1995, \pasp, 107, 966,
  \dodoi{10.1086/133645}

\bibitem[{{Van Der Walt} {et~al.}(2011){Van Der Walt}, {Colbert}, \&
  {Varoquaux}}]{vanderwalt2011}
{Van Der Walt}, S., {Colbert}, S.~C., \& {Varoquaux}, G. 2011, ArXiv e-prints.
\newblock \doarXiv{1102.1523}

\bibitem[{{Van Eylen} {et~al.}(2018){Van Eylen}, {Agentoft}, {Lundkvist},
  {Kjeldsen}, {Owen}, {Fulton}, {Petigura}, \& {Snellen}}]{VanEylen2018}
{Van Eylen}, V., {Agentoft}, C., {Lundkvist}, M.~S., {et~al.} 2018, \mnras,
  479, 4786, \dodoi{10.1093/mnras/sty1783}

\bibitem[{{Vogt} {et~al.}(2014){Vogt}, {Radovan}, {Kibrick}, {Butler},
  {Alcott}, {Allen}, {Arriagada}, {Bolte}, {Burt}, {Cabak}, {Chloros},
  {Cowley}, {Deich}, {Dupraw}, {Earthman}, {Epps}, {Faber}, {Fischer}, {Gates},
  {Hilyard}, {Holden}, {Johnston}, {Keiser}, {Kanto}, {Katsuki}, {Laiterman},
  {Lanclos}, {Laughlin}, {Lewis}, {Lockwood}, {Lynam}, {Marcy}, {McLean},
  {Miller}, {Misch}, {Peck}, {Pfister}, {Phillips}, {Rivera}, {Sandford},
  {Saylor}, {Stover}, {Thompson}, {Walp}, {Ward}, {Wareham}, {Wei}, \&
  {Wright}}]{Vogt2014}
{Vogt}, S.~S., {Radovan}, M., {Kibrick}, R., {et~al.} 2014, \pasp, 126, 359,
  \dodoi{10.1086/676120}

\bibitem[{{Wakeford} {et~al.}(2019){Wakeford}, {Wilson}, {Stevenson}, \&
  {Lewis}}]{Wakeford19}
{Wakeford}, H.~R., {Wilson}, T.~J., {Stevenson}, K.~B., \& {Lewis}, N.~K. 2019,
  Research Notes of the American Astronomical Society, 3, 7,
  \dodoi{10.3847/2515-5172/aafc63}

\bibitem[{{Weiss} \& {Marcy}(2014)}]{Weiss2014}
{Weiss}, L.~M., \& {Marcy}, G.~W. 2014, \apjl, 783, L6,
  \dodoi{10.1088/2041-8205/783/1/L6}

\bibitem[{{Weiss} {et~al.}(2018){Weiss}, {Marcy}, {Petigura}, {Fulton},
  {Howard}, {Winn}, {Isaacson}, {Morton}, {Hirsch}, {Sinukoff}, {Cumming},
  {Hebb}, \& {Cargile}}]{Weiss2018}
{Weiss}, L.~M., {Marcy}, G.~W., {Petigura}, E.~A., {et~al.} 2018, \aj, 155, 48,
  \dodoi{10.3847/1538-3881/aa9ff6}

\bibitem[{{Weiss} {et~al.}(2021){Weiss}, {Dai}, {Huber}, {Brewer}, {Collins},
  {Ciardi}, {Matthews}, {Ziegler}, {Howell}, {Batalha}, {Crossfield},
  {Dressing}, {Fulton}, {Howard}, {Isaacson}, {Kane}, {Petigura}, {Robertson},
  {Roy}, {Rubenzahl}, {Twicken}, {Claytor}, {Stassun}, {MacDougall}, {Chontos},
  {Giacalone}, {Dalba}, {Mocnik}, {Hill}, {Beard}, {Akana Murphy}, {Rosenthal},
  {Behmard}, {Van Zandt}, {Lubin}, {Kosiarek}, {Lund}, {Christiansen},
  {Matson}, {Beichman}, {Schlieder}, {Gonzales}, {Brice{\~n}o}, {Law}, {Mann},
  {Collins}, {Evans}, {Fukui}, {Jensen}, {Murgas}, {Narita}, {Palle},
  {Parviainen}, {Schwarz}, {Tan}, {Acton}, {Bryant}, {Chaushev}, {Gill},
  {Eigm{\"u}ller}, {Jenkins}, {Ricker}, {Seager}, \& {Winn}}]{Weiss2021}
{Weiss}, L.~M., {Dai}, F., {Huber}, D., {et~al.} 2021, \aj, 161, 56,
  \dodoi{10.3847/1538-3881/abd409}

\bibitem[{{Zechmeister} \& {K{\"u}rster}(2009)}]{ZechmeisterKurster}
{Zechmeister}, M., \& {K{\"u}rster}, M. 2009, \aap, 496, 577,
  \dodoi{10.1051/0004-6361:200811296}

\bibitem[{{Zeng} {et~al.}(2016){Zeng}, {Sasselov}, \& {Jacobsen}}]{Zeng2016}
{Zeng}, L., {Sasselov}, D.~D., \& {Jacobsen}, S.~B. 2016, \apj, 819, 127,
  \dodoi{10.3847/0004-637X/819/2/127}

\bibitem[{{Zhang} {et~al.}(2021){Zhang}, {Weiss}, {Huber}, {Blunt}, {Chontos},
  {Fulton}, {Grunblatt}, {Howard}, {Isaacson}, {Kosiarek}, {Petigura},
  {Rosenthal}, \& {Rubenzahl}}]{Zhang2021}
{Zhang}, J., {Weiss}, L.~M., {Huber}, D., {et~al.} 2021, arXiv e-prints,
  arXiv:2105.03446.
\newblock \doarXiv{2105.03446}

\bibitem[{{Zhu} \& {Wu}(2018)}]{Zhu2018}
{Zhu}, W., \& {Wu}, Y. 2018, \aj, 156, 92, \dodoi{10.3847/1538-3881/aad22a}

\end{thebibliography}

\section{Appendix}
\label{appendix}

\begin{table}[h!]
\footnotesize
\centering
 \caption{\textsc{Transit Mid-times}}
 \begin{tabular}{cccc}
 \hline
 \hline
Planet & Epoch \# & Mid-time (BJD) & Error (BJD)\\
\hline
b & 1 & 2458715.3552 & 0.0023 \\
b & 3 & 2458733.1156 & 0.0028 \\
b & 4 & 2458741.9962 & 0.0023 \\
b & 6 & 2458759.7587 & 0.0024 \\
b & 7 & 2458768.6376 & 0.0029 \\
b & 9 & 2458786.3987 & 0.0021 \\
b & 10 & 2458795.2825 & 0.0026 \\
b & 11 & 2458804.1602 & 0.0026 \\
b & 12 & 2458813.0424 & 0.0041 \\
b & 13 & 2458821.9181 & 0.0023 \\
b & 14 & 2458830.8014 & 0.0029 \\
b & 15 & 2458839.6806 & 0.0018 \\
b & 19 & 2458875.2017 & 0.0018 \\
b & 21 &  2458892.9626 & 0.0020 \\
b & 22 &  2458901.8430 & 0.0020 \\
b & 23 &  2458910.7222 & 0.0023 \\
b & 24 &  2458919.6017 & 0.0024 \\
b & 29 & 2458964.0028 & 0.0036 \\
b & 30 & 2458972.8837 & 0.0021 \\
b & 31 & 2458981.7649 & 0.0021 \\
b & 32 & 2458990.6450 & 0.0020 \\
b & 33 & 2458999.5248 & 0.0029 \\
b & 34 & 2459008.4056 & 0.0023 \\
\hline
c & 1 & 2458726.0546 & 0.0033 \\
c & 2 & 2458754.6340 & 0.0028 \\
c & 3 & 2458783.2116 & 0.0031 \\
c & 4 & 2458811.7992 & 0.0031 \\
c & 5 & 2458840.3758 & 0.0029 \\
c & 7 & 2458897.5359 & 0.0038 \\
c & 8 & 2458926.1168 & 0.0031 \\
\hline
d & 1 & 2458743.5531 & 0.0038 \\
d & 2 & 2458781.9029 & 0.0029 \\
d & 3 & 2458820.2645 & 0.0031 \\
d & 5 & 2458896.9613 & 0.0038 \\
d & 6 & 2458973.6648 & 0.0031 \\
\hline
\hline
\end{tabular} \\
\label{tbl:TTVs}
\end{table}

\begin{table}[h!]
\footnotesize
\centering
 \caption{\textsc{Radial Velocity Time Series}}
 \begin{tabular}{lccccc}
 \hline
 \hline
BJD & RV (m/s) & RV err (m/s) &  S-Value & S-Value err & Instrument \\
\hline
2458795.832 & -20.961 & 1.225 & 0.146 & 0.001 & HIRES \\
2458802.800 & -9.760 & 1.291 & 0.146 & 0.001 & HIRES \\
2458815.779 & -14.967 & 1.241 & 0.142 & 0.001 & HIRES \\
2458834.647 & 1.901 & 3.891 & 0.141 & 0.002 & APF \\
2458834.661 & 8.670 & 3.813 & 0.145 & 0.002 & APF \\
2458837.734 & -6.977 & 4.347 & 0.176 & 0.002 & APF \\
\hline
\hline
\end{tabular} \\
\footnotesize{{The full data set in a  machine readable format is available online.}}
\label{tbl:RVData}
\end{table}

\end{document}